\documentclass[traditabstract]{aa} % for the abstract without structuration 
\usepackage{graphicx}
\usepackage{txfonts}
\usepackage{natbib}
\bibpunct{(}{)}{;}{a}{}{,}

\newcommand{\Spitzer}{\textit{Spitzer}~}
\newcommand{\micron}{\rm \mu m}
\newcommand{\LX}{L_{\rm X}}
\newcommand{\Lx}{L_{\rm X}}
\newcommand{\Lv}{L_{\rm V}}
\newcommand{\Lm}{L_{\rm 12\,\micron}}
\newcommand{\Loiii}{L_{\rm [\ion{O}{iii}]}}

\newcommand{\logLv}{\log L_V}
\newcommand{\logLm}{\log L_{\rm 12\,\micron}}

\newcommand{\rsub}{r_\mathrm{sub}}
\newcommand{\ergs}{\rm erg/s}
\newcommand{\pc}{\rm pc}
\newcommand{\Mpc}{\rm Mpc}
\newcommand{\nh}{N_\mathrm{H}}
\newcommand{\LmLx}{L_\mathrm{MIR}-L_X}
\newcommand{\asec}{^{\prime\prime}}
\newcommand{\nFn}{\nu F_\nu}
\newcommand{\nuLnu}{\nu L_\nu}
\newcommand{\No}{N_0}

\begin{document}
   \title{The dusty heart of nearby active galaxies\thanks{Based on ESO observing programs 078.B-0303, 080.B-0240, 280.B-5068, 082.B-0299, and 083.B-0239.}}

   \subtitle{I. High-spatial resolution mid-IR spectro-photometry of Seyfert galaxies}
% \\ and its implication for the dust torus}

\titlerunning{The dusty heart of nearby active galaxies. I.}

\author{S.~F.~H\"onig\inst{1,2} \and
M.~Kishimoto\inst{1} \and
P.~Gandhi\inst{3} \and
A.~Smette\inst{4} \and
D.~Asmus\inst{4,5} \and
W.~Duschl\inst{5,6} \and
M.~Polletta\inst{7} \and
G.~Weigelt\inst{1}
}

\offprints{S.~F. H\"onig \\ \email{shoenig@physcis.ucsb.de}}

\institute{
Max-Planck-Institut f\"ur Radioastronomie, Auf dem H\"ugel 69, 53121 Bonn, Germany \and
University of California in Santa Barbara, Department of Physics, Broida Hall, Santa Barbara, CA 93109, USA \and
RIKEN Cosmic Radiation Lab, 2-1 Hirosawa, Wakoshi Saitama 351-0198, Japan \and
European Southern Observatory, Casilla 19001, Santiago 19, Chile \and
Institut f\"ur Theoretische Physik und Astrophysik, Christian-Albrechts-Universit\"at zu Kiel, Leibnizstr. 15, 24098 Kiel, Germany \and
Steward Observatory, The University of Arizona, 933 N. Cherry Ave, Tucson, AZ 85721, USA \and
INAF - IASF Milano, via E. Bassini, 20133 Milano, Italy}

   \date{Received November XX, 2009; accepted February 28, 2009}

% \abstract{}{}{}{}{} 
% 5 {} token are mandatory
 
\abstract{In a series of papers, we aim at stepping towards characterizing physical properties of the AGN dust torus by combining IR high-spatial resolution observations with 3D clumpy torus models. In this first paper, we present mid-IR imaging and $8-13\,\micron$ low-resolution spectroscopy of 9 type 1 and 10 type 2 AGN. The observations were carried out using the VLT/VISIR mid-IR imager and spectrograph and can be considered the largest currently available mid-infrared spectro-photometric data set of AGN at spatial resolution $\la$100\,pc. These data resolve scales at which the emission from the dust torus dominates the overall flux, and emission from the host galaxy (e.g. star-formation) is resolved out in most cases. The silicate absorption features are moderately deep and emission features, if seen at all, are shallow. The strongest silicate emission feature in our sample shows some notable shift of the central wavelength from the expected $9.7\,\micron$ (based on ISM extinction curves) to $\sim$10.5\,$\micron$. We compare the observed mid-IR luminosities of our objects to AGN luminosity tracers (X-ray, optical and [\ion{O}{iii}] luminosities) and found that the mid-IR radiation is emitted quite isotropically. In two cases, IC~5063 and MCG--3--34--64, we find evidence for extended dust emission in the narrow-line region. 
% It is speculated that outliers in the $\LmLx$-correlation might have extended dust emission in the narrow-line region (detected in MCG--3--34--64 and IC~5063). 
We confirm the correlation between observed silicate feature strength and Hydrogen column density which was recently found in \Spitzer data at lower spatial resolution. In a further step, our 3D clumpy torus model has been used to interpret the data. We show that the strength of the silicate feature and the mid-IR spectral index $\alpha$ can be used to get reasonable constraints on the radial dust distribution of the torus and the average number of clouds $\No$ along an equatorial line-of-sight in clumpy torus models. The mid-IR spectral index $\alpha$ is almost exclusively determined by the radial dust distribution power-law index $a$ and the silicate feature depth mostly depends on $\No$ and the torus inclination. A comparison of model predictions to our type 1 and type 2 AGN reveals that average parameters of $a=-1.0\pm0.5$ and $\No=5-8$ are typically seen in the presented sample, which means that the radial dust distribution is rather shallow. As a proof-of-concept of this method, we compared the model parameters derived from $\alpha$ and the silicate feature strength to more detailed studies of full IR SEDs and interferometry and found that the constraints on $a$ and $\No$ are consistent. Finally, we might have found evidence that the radial structure of the torus changes from low to high AGN luminosities towards steeper dust distributions, and we discuss implications for the IR size-luminosity relation.

}
   \maketitle
%
%________________________________________________________________

\section{Introduction}\label{sec:intro}

The dust torus is one of the key ingredients of the unification scheme of AGN \citep{Ant93,Urr95}. It must be pointed out, however, that in recent years the torus picture has evolved away from a ``donut'' towards a more general circumnuclear, geometrically- and optically-thick dust distribution. The dust reprocesses the optical/UV photons of the accretion disk and re-emits the received energy in the infrared (IR). Thus, the torus can be directly studied using the IR emission of AGN, unless the AGN is radio-loud. Within the last years, investigations based on IR SEDs characterized the torus emission and confirmed the basic picture of the unification scheme. Most notable are the silicate features at 10 and 18\,$\micron$ which are generally seen in absorption in type 2 AGN and in emission in type 1 objects.

However, more detailed studies -- mainly using data from the IRS and MIPS instruments on-board the \Spitzer satellite -- revealed subtle differences to model predictions. In particular, silicate emission features in type 1 AGN are much weaker than expected from early torus models which used smooth dust distributions. As shown in the literature, this might be explained by clumpiness of the dust in the torus which leads to an overall suppression of the emission feature \citep{Nen02,Dul05,Hon06,Sch08,Nen08b,Hon10}. However, as will be shown in the second paper of this series \citep[][, pre-print available on astro-ph]{Hon10}, weak emission features are not a generic property of clumpy torus models but might be used to constrain some model parameters of the dust distribution in clumpy tori. 
%A second surprise of the observed silicate emission features with \Spitzer was the detection of an apparent wavelength shift of the feature by $\sim0.5-1\,\micron$ towards longer wavelength.

\begin{table*}
\centering
\caption{Characteristics of our type 1 AGN mid-IR spectroscopic sample}\label{tab:scales}
\begin{tabular}{l c c c | c c c c | c c c }
\hline\hline
Object           & Type$^a$  & z      & $D_L$$^b$ & \multicolumn{4}{c|}{log Luminosities (erg/s)} & \multicolumn{3}{c}{Object scales}  \\
                 &           &        & (Mpc)     & $\Lx$ ($\nh$)\,$^c$ & $\Lv$\,$^d$ & $\Loiii$\,$^e$ & $\Lm$\,$^f$ & $\rsub$\,$^g$ (pc) & 8\,$\micron$ (pc)$^h$ & res. ($\rsub$) \\ \hline
         NGC~3227 &      S1.5 & 0.0049 & ~20.4     & 42.40 (22.80) & 42.20 & 40.51 & 42.63         & 0.014 & ~24.2 & 1680 \\
         NGC~3783 &      S1.5 & 0.0108 & ~44.7     & 43.19 (22.47) & 42.98 & 41.30 & 43.57         & 0.035 & ~52.5 & 1482 \\
         NGC~4593 &      S1.0 & 0.0101 & ~42.0     & 42.89 (20.30) & 42.88 & 40.69 & 43.18         & 0.032 & ~49.5 & 1560 \\
     ESO~323--G77 &      S1.2 & 0.0159 & ~66.3     & 42.76 (23.25) & 43.91 & 41.05 & 43.75         & 0.104 & ~77.1 & ~741 \\
   MCG--6--30--15 &      S1.5 & 0.0087 & ~35.8     & 42.86 (21.67) & 43.41 & 40.36 & 43.09         & 0.058 & ~42.3 & ~727 \\
         IC~4329A &      S1.2 & 0.0170 & ~70.5     & 43.87 (21.65) & 44.11 & 41.21 & 44.20         & 0.131 & ~82.0 & ~625 \\
         MARK~509 &      S1.5 & 0.0335 & 141~~     & 43.94 (20.70) & 44.01 & 42.29 & 44.25         & 0.116 & 158~~ & 1365 \\
         NGC~7213 &     S1.5L & 0.0051 & ~21.2     & 42.33 (20.60) & 42.76 & 39.87 & 42.50         & 0.028 & ~25.2 & ~913 \\
         NGC~7469 &      S1.5 & 0.0151 & ~62.7     & 43.33 (20.61) & 43.09 & 41.44 & 43.87         & 0.040 & ~73.1 & 1808 \\ \hline
\end{tabular}
\begin{list}{}{}
\item --- {\it Notes:} $^a$ AGN types from \citet{Ver06}: S: Seyfert, L: LINER; $^b$ Luminosity distance based on CMB reference frame redshifts from NED and $H_0=73$, $\Omega_\mathrm{m}=0.27$, and $\Omega_\mathrm{vac}=0.73$; $^c$ $\Lx$ = absorption-corrected 2$-$10\,keV luminosity from \citet{Dad07}, except for NGC~3227 \citep{Tue08}, logarithmic intrinsic Hydrogen column densities $\nh$ in units of cm$^{-2}$ from \citet{Tue08}, except ESO~323--G77 (average of \citet{Bec06} and \citet{Mal07}); $^d$ $\Lv\equiv\nuLnu$(5\,500\,$\AA$); from \citet{Ben09}, except NGC~7213 \citep{Win92a}, ESO~323--G77 \citep{Sch03}, and MCG--6--30--15 \citep{Ben06}. Additional extinction correction applied to IC~4329A and ESO~323--G77 \citep{Win92b}, and NGC~3783 \citep[$A_V=1.0$ from H$\alpha$/H$\beta$;][]{Gri92}; $^e$ [\ion{O}{iii}] total line fluxes from \citet{Stt03} (NGC~3783, NGC~4593, MCG--6--30--15, NGC~7213), \citet{Mel08} (NGC~3227, IC~4329A, MARK~509, NGC~7469), and \citet{Sch03} (ESO~323--G77); $^f$ $\Lm\equiv\nuLnu(12\,\micron)$; this work; $^g$from the fit to $K-$band reverberation mapping data as shown in \citet{Kis07}; observed $U-J$ or $U-K$ time lag radii: 0.071\,pc (NGC3783), 0.084\,pc (MARK509), 0.044\,pc (NGC7469), $^h$ Spatial scale corresponding to the theoretical diffraction limit for a 8.2\,m VLT telescope: 0$\farcs$25 at 8\,$\micron$, 0$\farcs$31 at 10\,$\micron$, 0$\farcs$37 at 12\,$\micron$; 
\end{list}
\end{table*}

One of the problems of \Spitzer is the comparably low spatial resolution of $\sim$3$\arcsec$ at 10\,$\micron$. On the other hand, the torus near- and mid-IR emission originates from the inner few parsecs around the AGN which corresponds to $<0\farcs1$ even for the nearest Seyfert galaxies. As a result, the \Spitzer data is prone to contamination from the host galaxy or circumnuclear star-formation which can be identified by strong PAH emission lines. One way to overcome this problem is a decomposition of the data into several components. However, this requires good knowledge about generic SEDs of each component, which is, of course, difficult when aiming at characterizing the torus emission, and introduces additional parameters which leads to parameter degeneracies.
% making physical interpretation less convincing.

The most direct way of studying the dust torus is IR interferometry. Recently, the IR emission source in a small number of nearby type 1 and type 2 AGN have been resolved in the mid-IR using the interferometric instrument MIDI at the VLTI \citep{Jaf04,Tri07,Bec08,Rab09,Tri09,Bur09} and in the near-IR using the Keck interferometer \citep{Swa03,Kis09}. While interferometry provides the most direct access to the torus structure and brightness distribution \citep{Kis09}, observations are limited to the brightest AGN which are in reach of current facilities. A compromise between \Spitzer and interferometry can be achieved by using ground-based 8\,m-class single telescopes. Such observations usually do not have the potential of firmly resolving the torus, but may isolate the nuclear dust emission from any contaminating source in the host galaxy \citep[e.g.][]{Hor09,Gan09,Mas09}. Following this approach, we observed 19 nearby AGN using the mid-infrared imager and spectrograph VISIR at the ESO Very Large Telescope (VLT) Paranal Observatory. In this second paper of our series, we present results from mid-IR spectro-photometric observations of our nearby AGN sample and interpret the observations with our 3D clumpy torus model \citep{Hon06,Hon10}. Details about the type 1 and type 2 sub-samples will be presented in Sect.~\ref{sec:sample}. The observations and data reduction are described in Sect.~\ref{sec:obs}. Sect.~\ref{sec:res} discusses the results of the observations. In Sect.~\ref{sec:plots}, we analyze the mid-IR characteristics of our sample and compare it to AGN properties. The data is then interpreted using our 3D clumpy torus model in Sect.~\ref{sec:model}. We summarize our main conclusions in Sect.~\ref{sec:summary}.

\section{Properties of the AGN sample}\label{sec:sample}

\subsection{Sample selection}

Our main goal is to obtain the highest spatial resolution spectro-photometric data set of AGN yet obtained in the mid-IR. However, the sample cannot be considered ``complete'' in any respect, and restrictions and selection criteria are outlined in the following. However, we consider the sample to be ``typical'' (or representative) for the Seyfert galaxy population in the Galactic vicinity (see below). Moreover, it should give us an idea of the ``clean'' nuclear emission at scales of tens of parsecs.

\begin{table*}
\caption{Characteristics of our type 2 AGN mid-IR spectroscopic sample}\label{tab:scales2}
\centering
\begin{tabular}{l c c c | c c c | c c c }
\hline\hline
Object           & Type$^a$  & z      & $D_L$$^b$ & \multicolumn{3}{c|}{log Luminosities (erg/s)} & \multicolumn{3}{c}{Object scales}  \\
                 &           &        & (Mpc)     & $\Lx$ ($\nh$)\,$^c$ & $\Loiii$\,$^d$ & $\Lm$\,$^e$ & $\rsub$\,$^f$ (pc) & 8\,$\micron$ (pc)$^g$ & res. ($\rsub$) \\ \hline
        NGC~2110 &       S1i & 0.0080 & ~33.2     & 42.60 (22.57) & 40.35 & 43.00         & \textit{0.028} & ~39.2 & \textit{1392} \\
    ESO~428--G14 &        S2 & 0.0063 & ~26.0     & $\ldots$ ($>$24) & 40.62 & 42.73         & \textit{0.020} & ~30.8 & \textit{1509} \\
    MCG--5-23-16 &       S1i & 0.0095 & ~39.5     & 43.24 (22.47) & 40.63 & 43.53         & \textit{0.052} & ~46.6 & \textit{~900} \\
        NGC~4507 &       S1h & 0.0128 & ~53.0     & 43.34 (23.46) & 41.57 & 43.69         & \textit{0.062} & ~62.1 & \textit{1004} \\
    MCG--3-34-64 &       S1h & 0.0176 & ~73.2     & 42.78 (23.69) & 41.45 & 44.14         & \textit{0.104} & ~84.9 & \textit{~814} \\
        NGC~5643 &        S2 & 0.0047 & ~19.2     & $\ldots$ ($>$24) & 40.55 & 42.49         & \textit{0.016} & ~22.9 & \textit{1471} \\
        NGC~5995 &      S1.9 & 0.0257 & 107~~     & 43.52 (21.95) & 42.01 & 44.06         & \textit{0.095} & 122~~ & \textit{1294} \\
         IC~5063 &       S1h & 0.0109 & ~45.3     & 42.87 (23.28) & 41.28 & 43.82         & \textit{0.072} & ~53.2 & \textit{~737} \\
        NGC~7582 &       S1i & 0.0044 & ~18.3     & 42.04 (22.98) & 40.15 & 42.71         & \textit{0.020} & ~21.8 & \textit{1093} \\
        NGC~7674 &       S1h & 0.0277 & 116~~     & 43.72 (24.00) & 41.93 & 44.26         & \textit{0.120} & 132~~ & \textit{1103} \\ \hline
\end{tabular}
\begin{list}{}{}
\item --- {\it Notes:} $^a$ AGN types from \citet{Ver06}: S: Seyfert, L: LINER, S1i: optical type 2 with broad Hydrogen lines in the infrared; S1h: type 2 with detected polarized optical broad lines; $^b$ Luminosity distance based on CMB reference frame redshifts from NED and $H_0=73$, $\Omega_\mathrm{m}=0.27$, and $\Omega_\mathrm{vac}=0.73$; $^c$ $\Lx$ = absorption-corrected 2$-$10\,keV luminosity from \citet{Dad07}, except IC~5063 \citep{Tue08}, MCG--3--34--64 and NGC~4507 \citep{Hon08}, NGC~5995 \citep{Hor08}, and NGC~7674 \citep{Bia05}; logarithmic intrinsic Hydrogen column densities in units of $10^{22}$\,cm$^{-2}$ from \citet{Tue08}, except ESO~428--G14 \citep{Del08}, NGC~5643 \citep{Mai98}, and NGC~7674 \citep[][changing-look AGN]{Bia05}; $^d$ [\ion{O}{iii}] fluxes from \citet{Stt03} (NGC~4507, IC~5063, NGC~7674), \citet{Mel08} (NGC~5643, NGC~7582), \citet{Gu06} (ESO~428--G14, MCG--3--34--64), \citet{Haa07} (NGC~2110), \citet{Win92a} (MCG--5--23--16), and \citet{Tra03} (NGC~5995); $^e$ $\Lm\equiv\nuLnu(12\,\micron)$; this work; $^f$ based on eq.~\ref{eq:sLm}); $^g$ Theoretical diffraction limit for a 8.2\,m VLT telescope: 0$\farcs$25 at 8\,$\micron$, 0$\farcs$31 at 10\,$\micron$, 0$\farcs$37 at 12\,$\micron$; 
\end{list}
\end{table*}

The original idea for selecting objects was based on a demand of high-spatial resolution. In particular, we aimed for objects which have a resolution $<$100\,pc around 10\,$\micron$ using the 8.2\,m UT3-telescope at Paranal. Thus, we are limited to AGN at angular-diameter distances $\le$70\,Mpc\,\footnote{Please note that the distances given in Table~\ref{tab:scales} \& \ref{tab:scales2} are {\it luminosity} distances $D_L$. The corresponding $D_L$ to our 70\,Mpc limit would be $D_L \le 73$\,Mpc.}. Since the ESO/Paranal observatory hosting VISIR is located in the Southern hemisphere, the whole sample is limited to objects mostly at Southern declinations. In addition, for easy execution of the observations in service mode, the AGN were selected to be brighter than 100\,mJy in most of the $N$-band. Based on these criteria, we mined the AGN catalog of \citet{Ver06} and compared the objects to previous low-spatial resolution mid-IR data from Spitzer and ISO to be able to estimate which objects are bright enough for VISIR. Since our ultimate goal concerns typical nearby AGN, i.e. Seyfert galaxies, all peculiar (e.g. LINERs) were skipped. To avoid synchrotron contamination, we also excluded radio-loud objects. Finally, since we are interested in the characteristics of the AGN, we avoided any objects where the nucleus is heavily obscured by host-galactic dust lanes, e.g. as in Circinus. Thus, any obscuration pattern seen in the data of most of our objects (i.e. silicate absorption features) is supposedly intrinsic to the nuclear environment. However, we acknowledge that IC~4329A and NGC~7582 have dust lanes passing over the nucleus, so that part of the observed absorption properties may originate in the host galaxy (in particular for the Seyfert 2 NGC~7582; see Sect.~\ref{sec:model} for possible consequences on our analysis and modeling).  In summary, we selected 9 type 1 and 10 type 2 AGN.

\subsection{Observed and intrinsic scales}

In Table~\ref{tab:scales} \& \ref{tab:scales2}, we list the basic characteristics of our type 1 and type 2 AGN sub-sample, respectively. In case the objects are Compton-thin, the intrinsic (absorption-corrected) 2$-$10\,keV X-ray luminosity can serve as a proxy for the total luminosity of the AGN. For the type 1 AGN (see Table~\ref{tab:scales}), we also provide optical luminosities which are more direct tracers for the accretion-disk luminosity. Observed scales for each objects are provided at the reference wavelength of $8\,\micron$.

Another way to look at our main requirement of high-spatial resolution is not the \textit{observed} scale but the \textit{intrinsic} scale of each object. Since most of the nuclear mid-IR radiation in radio-quiet Seyfert is presumably coming from dust emission, the fundamental scaling relation between the dust sublimation radius and the AGN luminosity, $\rsub \propto L^{1/2}$, provides such a distance-independent scaling. If we take UV/IR-reverberation mapping data \citep[e.g.][and references therein]{Sug06} and use the calibrated relation from \citet{Kis07}, we obtain $\rsub = 0.36\,\pc\,\times \left(\Lv/10^{45}\,\ergs\right)^{1/2}$. For the type 1 AGN, we can directly calculate the expected $\rsub$ from their optical luminosity and compare $\rsub$ to the spatial resolution of our observations (see Table~\ref{tab:scales}). As can be seen, all of the objects which meet our resolution selection $<100\,\pc$ (corresponding to $D_L\le73\,\Mpc$) have an \textit{intrinsic} spatial resolution $<2\,000\,\rsub$. Since the intrinsic scale is the ultimate factor determining how good the AGN can be isolated from the host galaxy, it is probably safe to include additional objects in the sample which have a spatial resolution $>$100\,pc, but still an intrinsic resolution $<2\,000\,\rsub$ without compromising our goal of highest spatial resolution. Thus, we include Markarian~509 which adds to the higher luminosity end of the type 1 sub-sample.

\begin{table*}
\caption{VISIR mid-IR photometry of our 9 type 1 AGN measured by Gaussian fitting to science target and calibrators. In case several calibrators were available, the fluxes and FWHM are mean values.}\label{tab:obs}
\centering
\begin{tabular}{l c c c c c c}
\hline\hline\
Object           &  Filter     & $\lambda_c$ & $\Delta\lambda$ & Flux             & $FWHM$              & Observing date \\
                 &  name       & ($\micron$) & ($\micron$)     & (mJy)            & target/calib        &                \\ \hline
        NGC~3227 &    ArIII    & ~8.99       & 0.14            &   $179.6\pm11.3$ & $0\farcs32/0\farcs31$ &2008-03-22T02:37\\
                 &    PAH2ref2 & 11.88       & 0.37            &   $320.1\pm21.9$ & $0\farcs37/0\farcs33$ &2008-03-22T02:41\\ \hline
%                 &    Q1       & 17.65       & 0.83            &  $853.1\pm215.0$ & $0\farcs52/0\farcs46$ &2008-03-22T02:22\\ \hline
        NGC~3783 &    PAH1     & ~8.59       & 0.42            &   $323.9\pm33.7$ & $0\farcs31/0\farcs26$ &2008-03-20T06:08\\
                 &    ArIII    & ~8.99       & 0.14            &   $350.9\pm23.2$ & $0\farcs30/0\farcs29$ &2008-03-20T06:16\\
                 &    SIV      & 10.49       & 0.16            &   $558.4\pm23.2$ & $0\farcs31/0\farcs31$ &2005-04-17T01:27\\
                 &    PAH2ref2 & 11.88       & 0.37            &   $595.8\pm39.2$ & $0\farcs34/0\farcs32$ &2008-03-20T06:20\\
                 &    NeIIref1 & 12.27       & 0.18            &   $685.0\pm37.2$ & $0\farcs35/0\farcs32$ &2005-04-17T01:36\\ \hline
%                 &    Q1       & 17.65       & 0.83            & $1466.3\pm198.6$ & $0\farcs48/0\farcs46$ &2008-03-20T06:11\\
%                 &    Q2       & 18.72       & 0.88            & $1660.2\pm312.9$ & $0\farcs51/0\farcs50$ &2006-03-13T05:51\\ \hline
        NGC~4593 &    ArIII    & ~8.99       & 0.14            &   $181.8\pm13.8$ & $0\farcs31/0\farcs26$ &2008-04-01T06:32\\
                 &    PAH2ref2 & 11.88       & 0.37            &   $278.8\pm20.4$ & $0\farcs36/0\farcs31$ &2008-04-01T06:35\\ \hline
%                 &    Q1       & 17.65       & 0.83            &   $425.9\pm32.8$ & $0\farcs49/0\farcs44$ &2008-04-01T06:16\\ \hline
    ESO~323--G77 &    ArIII    & ~8.99       & 0.14            &   $298.9\pm11.5$ & $0\farcs40/0\farcs29$ &2009-05-19T00:58\\
                 &    PAH2ref2 & 11.88       & 0.37            &    $377.4\pm7.6$ & $0\farcs37/0\farcs36$ &2009-05-10T01:01\\ \hline
  MCG--6--30--15 &    SIV      & 10.49       & 0.16            &   $311.7\pm11.4$ & $0\farcs32/0\farcs38$ &2006-04-14T03:58\\
                 &    PAH2     & 11.25       & 0.59            &   $359.6\pm24.7$ & $0\farcs33/0\farcs33$ &2006-04-14T04:03\\
                 &    NeIIref1 & 12.27       & 0.18            &   $377.5\pm11.2$ & $0\farcs36/0\farcs36$ &2006-04-14T04:07\\ \hline
        IC~4329A &    ArIII    & ~8.99       & 0.14            &   $743.4\pm22.0$ & $0\farcs30/0\farcs29$ &2009-05-19T01:12\\
                 &    PAH2ref2 & 11.88       & 0.37            &  $1014.1\pm18.8$ & $0\farcs36/0\farcs36$ &2009-05-10T01:15\\ \hline
        MARK~509 &        SIV  & 10.49       & 0.19            &   $207.3\pm17.0$ & $0\farcs35/0\farcs34$ &2006-06-14T10:20\\
                 &        PAH2 & 11.25       & 0.59            &   $240.8\pm27.1$ & $0\farcs32/0\farcs32$ &2006-06-14T10:25\\
                 &        NeII & 12.81       & 0.21            &   $243.0\pm19.8$ & $0\farcs30/0\farcs35$ &2006-06-14T10:30\\ \hline
        NGC~7213 &    SIV      & 10.49       & 0.16            &   $239.1\pm22.0$ & $0\farcs33/0\farcs32$ &2006-07-14T09:58\\
                 &    PAH2     & 11.25       & 0.59            &   $273.3\pm34.5$ & $0\farcs37/0\farcs34$ &2006-07-14T10:03\\
                 &    NeIIref1 & 12.27       & 0.18            &   $245.6\pm16.9$ & $0\farcs35/0\farcs32$ &2006-07-14T10:07\\ \hline
        NGC~7469 &    SIV      & 10.49       & 0.16            &   $447.8\pm15.6$ & $0\farcs39/0\farcs31$ &2006-07-12T07:21\\
                 &    PAH2     & 11.25       & 0.59            &   $470.1\pm38.8$ & $0\farcs42/0\farcs34$ &2006-07-12T07:25\\
                 &    PAH2ref2 & 11.88       & 0.37            &   $505.9\pm25.4$ & $0\farcs34/0\farcs32$ &2006-07-14T09:29\\
                 &    NeIIref1 & 12.27       & 0.18            &   $595.0\pm18.0$ & $0\farcs36/0\farcs34$ &2006-07-12T07:30\\
                 &    NeIIref2 & 13.04       & 0.22            &   $629.7\pm16.6$ & $0\farcs35/0\farcs47$ &2008-08-26T06:48\\ \hline
%                 &    Q1       & 17.65       & 0.83            &  $992.9\pm123.3$ & $0\farcs51/0\farcs49$ &2008-10-02T00:43\\
%                 &    Q3       & 19.50       & 0.40            & $1257.7\pm160.0$ & $0\farcs52/0\farcs53$ &2008-10-02T00:58\\ \hline
\end{tabular}
\begin{list}{}{}
\item 
\end{list}
\end{table*}

Since the UV/optical emission from the accretion disk is thought to be obscured in type 2 AGN, a direct comparison between observed scales and intrinsic scales is not possible via the reverberation-based size-luminosity relation. However, using our type 1 sub-sample, we are able to convert the size-luminosity relation from the optical to other wavebands. Since we are dealing with mid-IR observations, the most convenient way is a conversion of the size-luminosity relation to 12\,$\micron$. This wavelength is mostly outside the silicate feature so that it is not affected by possible anisotropies due to silicate absorption or emission. Recent high-spatial resolution studies (also using the VISIR instrument on the 8.2-m VLT/UT3 telescope) have shown that type 1 and type 2 AGN follow basically the same $\LmLx$-relation in the luminosity range which is covered by our sample \citep{Hor08,Gan09}. Anisotropy between the two samples are within the observational errors and are smaller than about a factor of 2 to 3. Thus, $\Lm$ can also serve as a proxy for the AGN luminosity of our objects and enables us to estimate intrinsic scales. We note that this is only a good approximation as long as high spatial resolution data is available.

As a first step, we determine the correlation between $L_V$ and $\Lm$ in the type 1 sub-sample, assuming that the covering factor is similar for all objects. We find that $\logLv = (6.9\pm9.6)+(0.84\pm0.22)\times\logLm$ (Spearman rank 0.88, null-hypothesis probability $5\times10^{-3}$). Within errors, this is consistent with $L_V \propto \Lm$, with a ratio $L_V/\Lm = 0.59^{+0.82}_{-0.34}$, which we assume in the following. Using this correlation, we obtain a scaling relation for our AGN sample of
\begin{equation}
\rsub = (0.28^{+0.15}_{-0.10})\,\pc\,\times\left(\frac{\Lm}{10^{45}\,\ergs}\right)^{1/2} \label{eq:sLm}
\end{equation}
Despite the uncertainty of a factor of 1.5, this should enable us to give at least an estimate of intrinsic scales of our type 2 objects. Intrinsic scales based on eq.~(\ref{eq:sLm}) are given in Table~\ref{tab:scales2}. As can be seen, while NGC~5995 and NGC~7674 have observed scales $>$100\,pc, their intrinsic scales are $<2\,000\,\rsub$, even when accounting for the factor of 1.5 uncertainty in eq.~(\ref{eq:sLm}). These two objects have been added to our sample to have higher luminosity Seyferts in both the type 1 and type 2 sub-samples. We note that the relation based on $\Lx$ instead of $\Lm$ would not improve the results ($\rsub = 0.45^{+0.32}_{-0.18}\,\pc\,\times\left(\Lx/10^{45}\,\ergs\right)^{1/2}$).

\begin{table*}
\caption{VISIR mid-IR photometry of our 10 type 2 AGN measured by Gaussian fitting to science target and calibrators. In case several calibrators were available, the fluxes and FWHM are mean values.}\label{tab:obs2}
\centering
\begin{tabular}{l c c c c c c}
\hline\hline\
Object           &  Filter     & $\lambda_c$ & $\Delta\lambda$ & Flux             & $FWHM$              & Observing date \\
                 &  name       & ($\micron$) & ($\micron$)     & (mJy)            & target/calib        &                \\ \hline
        NGC~2110 &    PAH1     & ~8.59       & 0.42            &    $156.6\pm9.9$ & $0\farcs33/0\farcs32$ &2006-10-10T07:28\\
                 &    ArIII    & ~8.99       & 0.14            &    $169.1\pm6.4$ & $0\farcs31/0\farcs29$ &2006-10-10T07:31\\
                 &    PAH2ref2 & 11.88       & 0.37            &    $294.3\pm8.0$ & $0\farcs33/0\farcs35$ &2006-10-10T07:40\\ \hline
%                 &    Q1       & 17.65       & 0.83            &   $555.1\pm72.9$ & $0\farcs48/0\farcs47$ &2006-10-10T07:50\\ \hline
    ESO~428--G14 &    PAH1     & ~8.59       & 0.42            &   $112.7\pm18.5$ & $0\farcs37/0\farcs41$ &2007-03-01T03:27\\
                 &    ArIII    & ~8.99       & 0.14            &   $122.8\pm14.4$ & $0\farcs40/0\farcs30$ &2007-03-01T03:34\\
                 &    PAH2ref2 & 11.88       & 0.37            &    $207.1\pm9.4$ & $0\farcs43/0\farcs32$ &2007-03-01T03:42\\ \hline
%                 &    Q1       & 17.65       & 0.83            &   $736.9\pm20.6$ & $0\farcs69/0\farcs46$ &2007-03-01T03:50\\ \hline
  MCG--3--34--64 &    SIV      & 10.49       & 0.16            &   $591.9\pm22.4$ & $0\farcs39/0\farcs25$ &2006-04-09T05:23\\
                 &    PAH2     & 11.25       & 0.59            &   $685.8\pm51.0$ & $0\farcs43/0\farcs31$ &2006-04-09T05:31\\
                 &    NeIIref1 & 12.27       & 0.18            &   $889.3\pm21.2$ & $0\farcs44/0\farcs34$ &2006-04-09T05:35\\
                 &    NeIIref2 & 13.04       & 0.22            &  $1116.0\pm60.7$ & $0\farcs46/0\farcs36$ &2006-01-18T07:58\\ \hline
        NGC~4507 &    SIV      & 10.49       & 0.16            &   $489.2\pm43.3$ & $0\farcs32/0\farcs32$ &2006-04-15T02:58\\
                 &    PAH2     & 11.25       & 0.59            &   $554.7\pm43.6$ & $0\farcs32/0\farcs33$ &2006-04-15T03:02\\
                 &    NeIIref1 & 12.27       & 0.18            &   $631.4\pm10.8$ & $0\farcs34/0\farcs34$ &2006-04-15T03:07\\ \hline
  MCG--5--23--16 &    PAH1     & ~8.59       & 0.42            &   $337.2\pm10.5$ & $0\farcs27/0\farcs33$ &2007-01-30T06:09\\
                 &    ArIII    & ~8.99       & 0.14            &   $358.2\pm16.7$ & $0\farcs28/0\farcs32$ &2007-01-30T06:32\\
                 &    PAH2ref2 & 11.88       & 0.37            &   $633.4\pm24.0$ & $0\farcs33/0\farcs35$ &2007-01-30T06:40\\ \hline
%                 &    Q1       & 17.65       & 0.83            & $1437.5\pm179.3$ & $0\farcs46/0\farcs46$ &2007-01-30T06:24\\
%                 &    Q2       & 18.72       & 0.88            &  $1488.7\pm71.5$ & $0\farcs49/0\farcs48$ &2006-03-13T05:12\\ \hline
        NGC~5643 &    ArIII    & ~8.99       & 0.14            &   $143.3\pm33.5$ & $0\farcs35/0\farcs27$ &2009-05-03T04:58\\
                 &    PAH2ref2 & 11.88       & 0.37            &    $287.1\pm9.1$ & $0\farcs37/0\farcs35$ &2009-05-03T05:01\\ \hline
        NGC~5995 &        SIV  & 10.49       & 0.16            &   $266.0\pm10.7$ & $0\farcs37/0\farcs38$ &2006-04-14T04:48\\
                 &        PAH2 & 11.25       & 0.59            &   $332.5\pm17.7$ & $0\farcs38/0\farcs31$ &2006-04-14T04:52\\
                 &        NeII & 12.81       & 0.21            &   $381.8\pm20.2$ & $0\farcs40/0\farcs33$ &2006-04-14T04:57\\ \hline
         IC~5063 &    SIV      & 10.49       & 0.16            &   $608.6\pm22.0$ & $0\farcs47/0\farcs37$ &2006-05-05T09:54\\
                 &    PAH2     & 11.25       & 0.59            &   $726.6\pm24.7$ & $0\farcs43/0\farcs38$ &2006-05-05T09:59\\
                 &    PAH2ref2 & 11.88       & 0.37            &   $924.9\pm25.0$ & $0\farcs42/0\farcs38$ &2005-06-10T08:06\\
                 &    NeIIref1 & 12.27       & 0.18            &  $1036.5\pm56.2$ & $0\farcs43/0\farcs37$ &2006-05-05T10:03\\ \hline
        NGC~7582 &    PAH1     & ~8.59       & 0.42            &   $327.2\pm37.7$ & $0\farcs51/0\farcs41$ &2008-08-18T09:29\\
                 &    ArIII    & ~8.99       & 0.14            &   $235.9\pm27.2$ & $0\farcs33/0\farcs29$ &2008-08-23T08:41\\
                 &    PAH2ref2 & 11.88       & 0.37            &   $383.6\pm26.2$ & $0\farcs38/0\farcs33$ &2008-08-23T08:38\\ \hline
         NGC7674 &        NeII & 12.81       & 0.21            &   $518.4\pm21.7$ & $0\farcs45/0\farcs38$ &2006-07-13T08:55\\ \hline
\end{tabular}
\begin{list}{}{}
\item 
\end{list}
\end{table*}

In summary, our objects can be described as a sample of typical nearby Seyfert galaxies -- obscured and non-obscured -- spanning the luminosity range from about $10^{42}\,\ergs$ to $10^{44}\,\ergs$. The intrinsic spatial resolution achieved by our observations is $<2\,000\,\rsub$. Please note that we accidentally included one optically broad-line (BL) LINER in our sample, NGC~7213 (allegedly a ``type 1 LINER''), which will be treated as a type 1 AGN in the rest of the paper but discussed separately where applicable.

\section{Observations and data reduction}\label{sec:obs}

\subsection{Spectroscopy}

\begin{figure}
\centering
\includegraphics[width=0.5\textwidth]{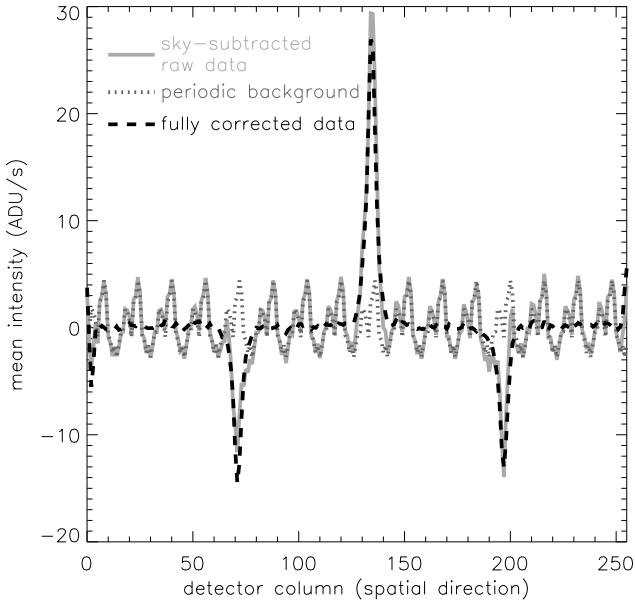}
\caption{Illustration of the periodic background map (PBM) removal for the $12.4\,\micron$ setting of NGC~7213. The solid light-gray line represents the raw data after sky subtraction (mean over all rows). The PBM is shown by the dotted dark-gray line. The dashed black line presents the final data after PBM removal.}\label{fig:pbg}
\end{figure}

We used the VISIR mid-infrared imager and spectrograph mounted on the 8.2\,m UT3 telescope at the ESO/Paranal observatory in Chile. The observations have been carried out in service mode in ESO periods 78, 80, 82, and 83. In total, 9 Seyfert 1 and 10 Seyfert 2 galaxies have been observed in low-spectral (LR) resolution mode ($R\sim300$). The standard configuration for LR long-slit spectroscopy results in a pixel resolution of 0$\farcs$127, which samples the PSF in the $N$-band by $\sim$2.5 pixels. To cover the full $N$-band, 4 different spectral settings have to be used with central wavelengths at 8.5, 9.8, 11.4, and 12.4\,$\micron$. We used a slit width of 0$\farcs$75 which is approximately 2$-$3 times larger than the FWHM achieved with VISIR ($0\farcs25-0\farcs39$ across the $N$-band) and minimizes the risk of slit losses. All of our targets are unresolved point-sources, except for NGC~7469 where some off-nuclear emission is expected from the circumnuclear star-burst ring.

For the initial steps of data reduction, we followed the standard ESO pipeline. The individual chopped nodding cycles (chopping width 8$\arcsec$, nodding in parallel mode) were combined and wavelength-calibrated using the Common Pipeline Library (CPL) recipes in {\it ESOREX}. After that, we used our own procedures to calibrate and extract the spectra. First, the remaining sky-offset has been removed. For that, we fitted a low-order polynomial in spatial direction to each wavelength bin (typically of orders 0 to 2). We note that this procedure can only be used because we are not interested in very large-scaled smooth structure --- which would not be visible anyway because of the flux limit. This polynomial sky subtraction flattened the image background significantly (see Fig.~\ref{fig:pbg}). 

However, a periodic background pattern in spatial direction remained in each wavelength bin, with a frequency length of 16 pixels and not depending on chopping width, frequency, or position angle. In usual observing conditions, this pattern dominates the total variance of the background. 
%While it is only a few ADU/s in amplitude and, thus, does not affect bright point-like targets too much, it is very disturbing when dealing with faint mid-IR objects such as AGN. 
The ratio of object peak to background variation peak can be as low as 4:1 in AGN, thus being very disturbing when dealing with faint mid-IR objects. We developed an efficient method to remove this background pattern using a ``periodic background map (PBM)''. The PBM is generated by creating a cube with several copies of the sky-removed science array, each copy shifted by 16 pixels in spatial direction with respect to the previous copy. Finally, all shifted science-array copies are combined by median-filtering each pixel. The resulting PBM contains only the 16-pixel-frequency background without flux from the (point-like or slightly resolved) science target. In Fig.~\ref{fig:pbg}, we illustrate the PBM removal for the $12.4\,\micron$ setting of NGC~7213. The solid light-gray line represents the raw data after sky subtraction (mean over all rows). For these data, the PBM has been determined, as shown by the dotted dark-gray line. The dashed black line shows the final data after PBM removal. As can be seen, the noise variation is significantly suppressed as compared to the original data.

The described procedure has been applied to both science and calibrator data. After removing all background, wavelength-dependent conversion factors have been determined from the calibrators and the science data were flux-calibrated accordingly. We refrained from airmass corrections since the differential airmasses between target and calibrator were rather small, so that corrections would be within the calibration errors. In the case of MCG--3--34--64, we found a beam-centering problem. In fact, visual inspection of acquisition images revealed that only part of the object was placed within the 0$\farcs$75 slit. We used the acquisition images in the N\_SW filter (made through the open slit) of the science target and the calibrator to quantify the loss due to this problem. The measured flux for MCG--3--34--64 in this filter is $F(\mathrm{N\_SW})=421.06\pm24.75$\,mJy (wavelength: $8.85\pm1.35\,\micron$). Comparing to the integrated flux in the spectrum, we found that only about $70\pm4$\% of the object flux was located within the slit. Thus, we corrected the flux-calibrated spectra of MCG--3--34--64 accordingly.

\subsection{Imaging photometry}

%\subsubsection{Mid-IR photometry}

\begin{figure*}
%\centering
\includegraphics[width=1.0\textwidth]{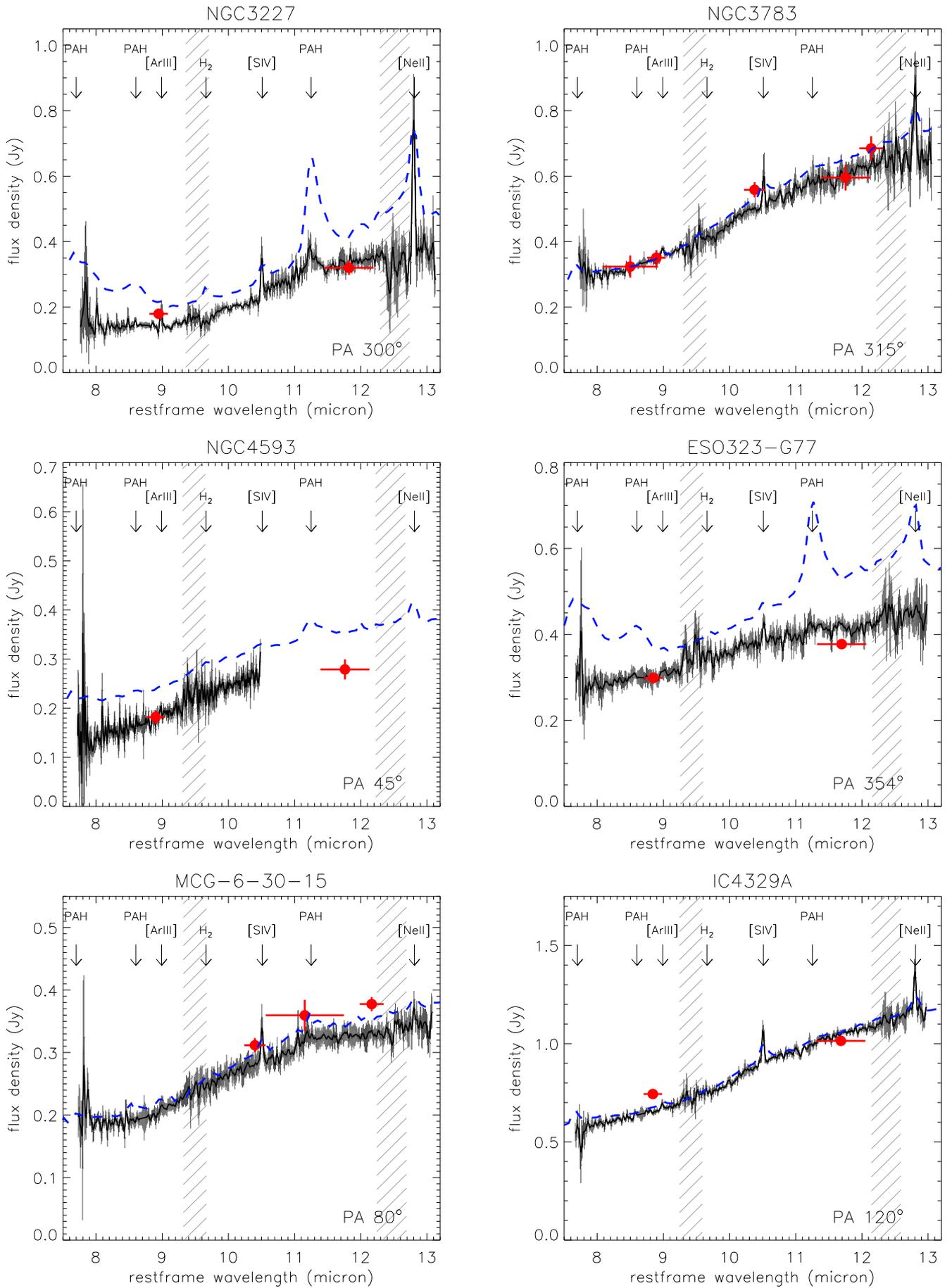}
\caption{VISIR low-resolution spectroscopy (black solid line and gray error bars) and photometry (red filled circles) of 8 type 1 AGN. Overplotted (blue dashed line) are Spitzer data with approximately 10 times less spatial resolution. The hatched areas mark regions with strong sky lines which are difficult to calibrate. Prominent mid-IR emission line positions are indicated. The slit position angle (PA) is given in the lower right corner of each panel.}\label{fig:VISIR_type1}
\end{figure*}

\begin{figure*}
%\ContinuedFloat
\addtocounter{figure}{-1}
%\centering
\includegraphics[width=1.0\textwidth]{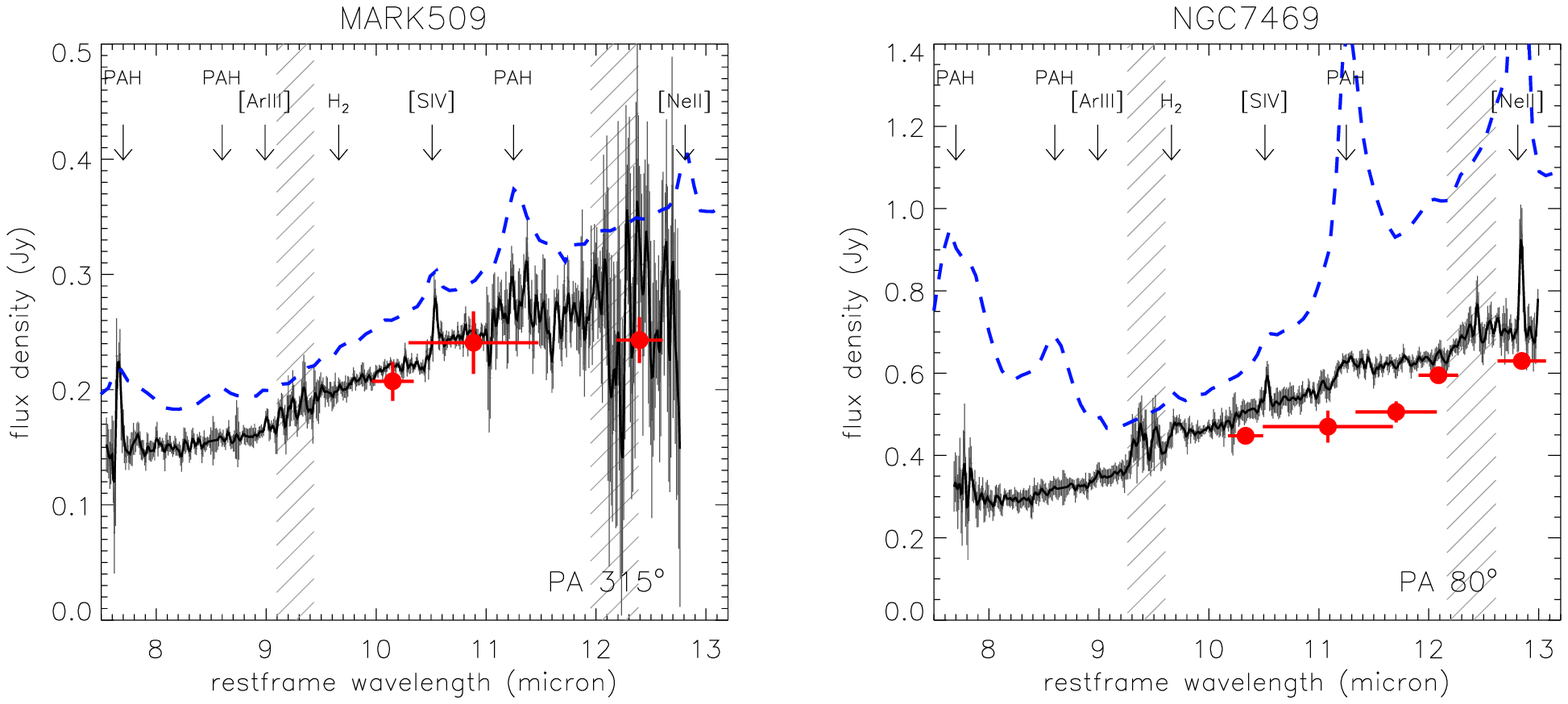}
\caption{--- {\it continued.}}
\end{figure*}

Along with the spectroscopy data, we acquired VISIR images in several mid-infrared filters to compare and validate the absolute and relative flux calibration of our spectra. Together with archival VISIR data, all of our objects were observed in at least 2 different narrow-band filters between 8 and 13\,$\micron$, except NGC~7674 (only 1 filter). For photometry, the standard CPL pipeline products were used for both science target and calibrator. The imaging was performed in parallel or perpendicular chop/nod mode, so that 3 or 4 beams appear on the detector for the science data (parallel: 2 negative single-beams and 1 positive double-beam). On the other hand, the standard stars were observed almost exclusively in perpendicular chop/nod mode resulting in 4 single-beams (2 negative, 2 positive beams). 

Each science and calibrator beam was fitted by a 2-dimensional Gaussian to resemble the core of the Airy disk of the PSF. We confirmed that our science targets are point sources in the mid-IR by comparing the Gaussian width of the targets with the standard stars. The measured Gaussian FWHM of science target and associated calibrator\footnote{Where several calibrators were available, the given calibrator FWHM is the average of all calibrators.} are given in Table~\ref{tab:obs}. Although the science targets usually have slightly larger FWHM than the calibrators, the differences are not significant, except for MCG--3--34--64 and IC~5063 which we will discuss in Sect.~\ref{sec:lmircont} in detail. In most cases, the differences between the FWHM of the science targets and calibrators are probably the result of PSF instabilities of VISIR which are known to be a common problem \citep{Hor09}. Therefore we consider that our sources are mostly unresolved. After the Gaussians have been removed, all science targets showed the first Airy ring without any additional emission source, except for NGC~7469. This galaxy has a well-known nuclear star-burst ring within 2$\arcsec$ of the nucleus which we spatially resolve in our observations and which contains more flux than the first Airy ring.

\begin{figure}
%\centering
\includegraphics[width=0.5\textwidth]{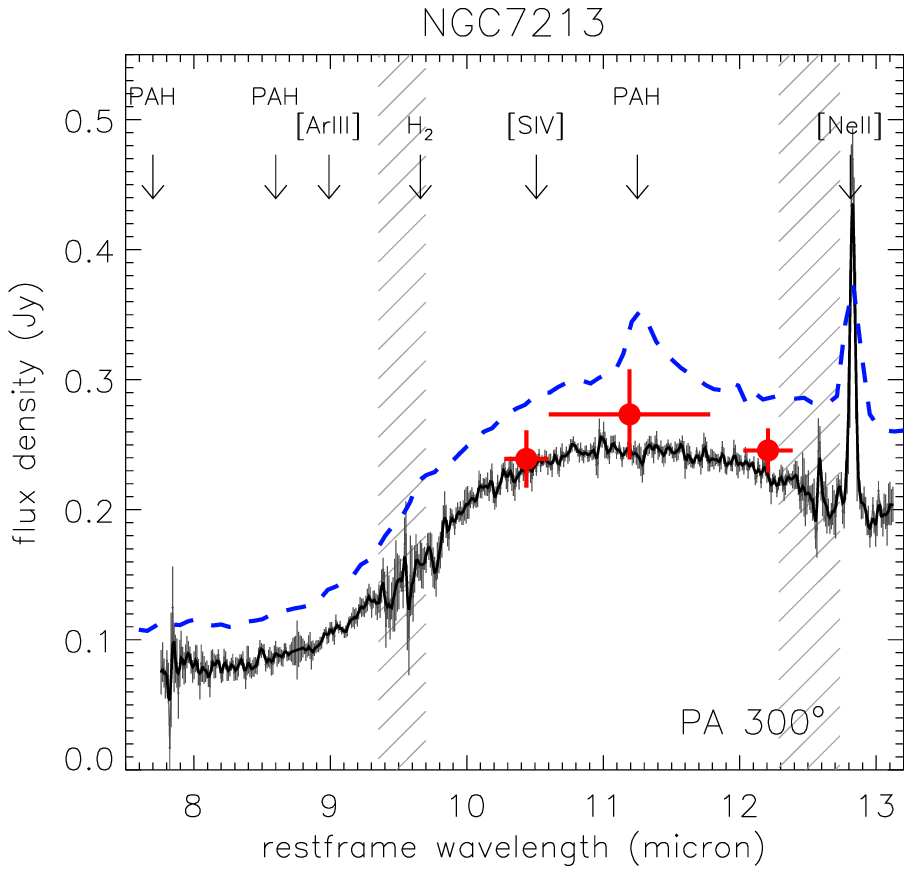}
\caption{VISIR low-resolution spectroscopy (black solid line and gray error bars) and photometry (red filled circles) of the LINER galaxy NGC~7213. Overplotted (blue dashed line) are Spitzer data with approximately 10 times less spatial resolution. The hatched areas mark regions with strong sky lines which are difficult to calibrate. Prominent mid-IR emission line positions are indicated. The slit position angle (PA) is given in the lower right corner.}\label{fig:VISIR_type1_L}
\end{figure}

The final fluxes were obtained by calculating conversion factors from the integrated intensity of the Gaussian fits of the calibrators and using these factors for the respective integrated Gaussian intensity of the science targets. Since we are only interested in point source fluxes, this method of optimized extraction has the advantage that any flux from off-nuclear emission is excluded. This provides the best possible estimate for the torus emission which is presumably unresolved. The resulting fluxes are presented in Tables~\ref{tab:obs} \& \ref{tab:obs2}. All data was obtained with a chop throw of 8\arcsec, except for the archival PAH2ref2 filter data of NGC~7469 and the archival NeIIref2 filter data of MCG--3--34--64 where a chop throw of 10\arcsec was used.

%\subsubsection{Near-IR photometry}
%
%In addition to the VISIR data, we used archival NACO and ISAAC imaging data to extend the SED to shorter wavelengths where such data was available. There are, however, several caveats by comparing the archival near-IR photometry to our mid-IR observations. First, in the near-IR $J$-, $H$-, and $K$-bands, the host galactic bulge can contribute significantly to the nuclear emission which potentially contaminates our torus SEDs. Detailed host decomposition might help extracting nuclear near-IR fluxes. However, in the case of NACO different AO preformance for target and calibrator may affect the photometric accuracy \citep[e.g. as illustrated for NGC~1068 in][]{Wei04}. Second, type 1 AGN can show a substantial variability in the near-IR \citep{Gla92,Gla04}. Typical $K$-band 1$-\sigma$ variations range from 0.2 to 1\,mag, which means a change of a factor of 1.2$-$2.5 in flux. Since the variations occur on a timescale of months to years, they can severly affect SEDs consisting of data from different epochs.
%
%To overcome, at least, the host-galaxy contamination, we subtracted the median background at 1$\farcs$3 from each pixel and fitted a Gaussian to the central emission peak in both target and calibrator frames. While there might still be some remaining host flux in the PSF, this should affect mostly our $K$-band data of NGC~3783 while longer wavelength ($L$- and $M$-bands) should be less affected by host emission in general. The resulting fluxes are shown in Table~\ref{tab:obs}.

\section{Results and discussion}\label{sec:res}

\subsection{VISIR spectra and photometry}

\begin{figure*}
%\centering
\includegraphics[width=1.0\textwidth]{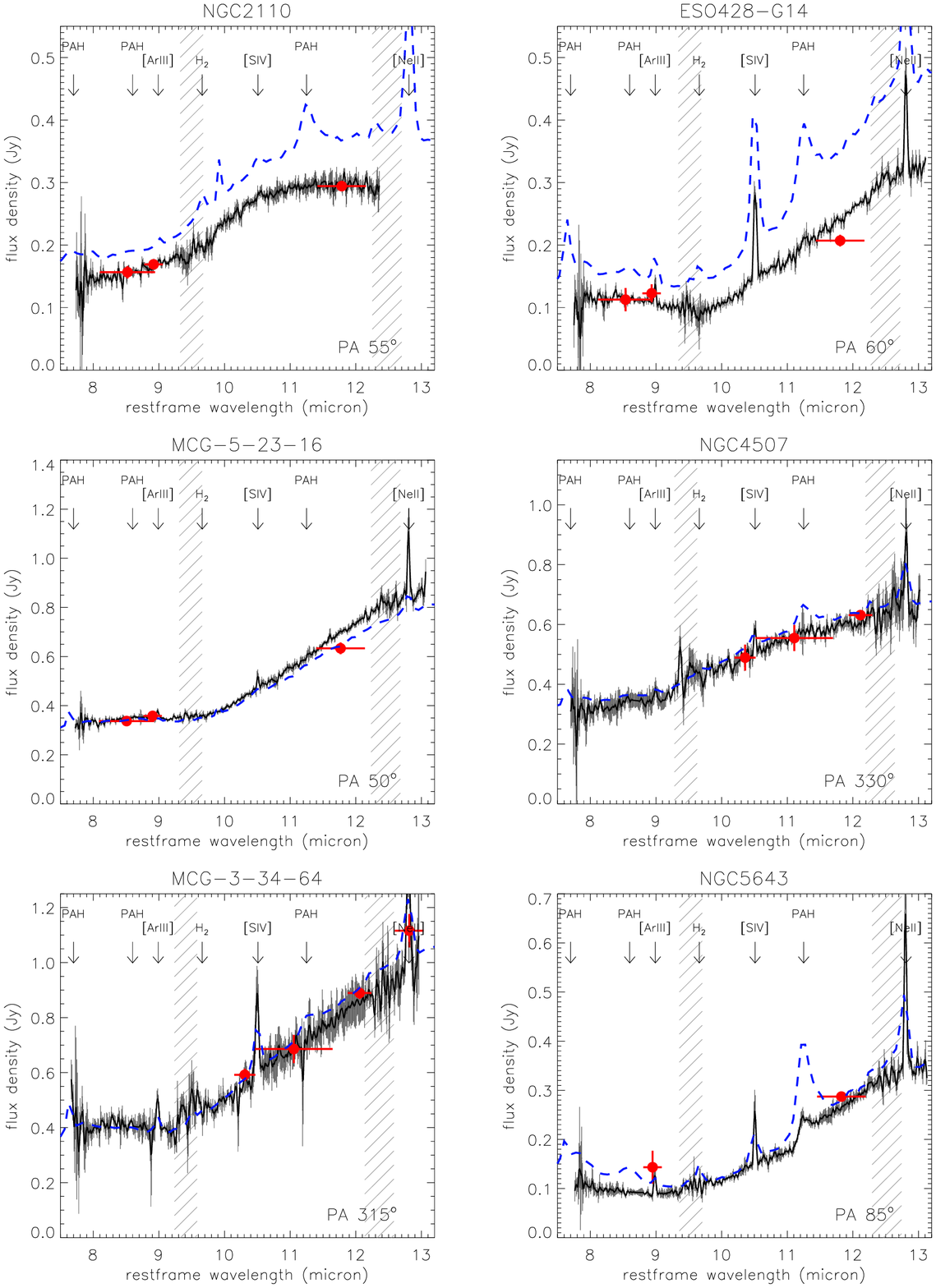}
\caption{VISIR low-resolution spectroscopy (black solid line and gray error bars) and photometry (red filled circles) of 10 type 2 AGN. Overplotted (blue dashed line) are Spitzer data with approximately 10 times less spatial resolution. The hatched areas mark regions with strong sky lines which are difficult to calibrate. Prominent mid-IR emission line positions are indicated. The slit position angle (PA) is given in the lower right corner of each panel.}\label{fig:VISIR_type2}
\end{figure*}

\begin{figure*}
%\ContinuedFloat
%\centering
\addtocounter{figure}{-1}
\includegraphics[width=1.0\textwidth]{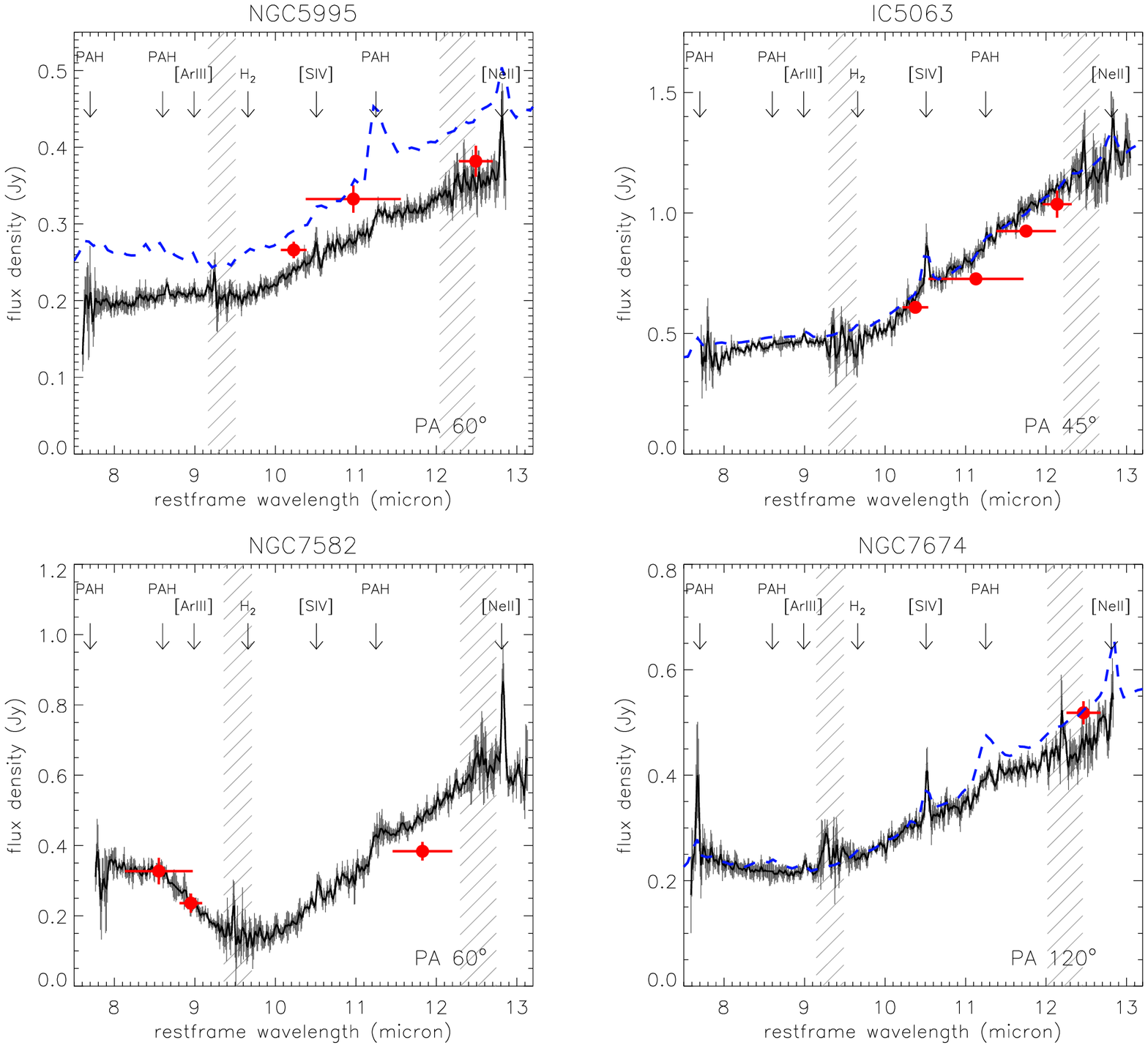}
\caption{--- {\it continued.}}
\end{figure*}

In Figs.~\ref{fig:VISIR_type1} \& \ref{fig:VISIR_type1_L} we show the individual VISIR $8-13\,\micron$ spectra of our type 1 AGN (8 Seyfert galaxies and 1 BL LINER) together with results from VISIR photometry. In the lower-right corner of each panel, we noted the slit position angle (PA) for each object. The corresponding data of the 10 Seyfert 2 galaxies are shown in Fig.~\ref{fig:VISIR_type2}. In addition to these data, the corresponding \Spitzer IRS low-resolution spectra are shown for the same wavelength range. The spatial resolution of the IRS spectra is $\sim3\arcsec$ while the VISIR data resolves scales of $\sim0\farcs3$ at $10\,\micron$. For NGC~7582, no \Spitzer spectrum was available yet. The positions of prominent emission lines seen in AGN are shown. The gray-hatched areas in each spectrum mark regions of extensive sky-line emission/absorption which are difficult to calibrate due to some degree of variability within the night and/or compromised S/N ratios. 

All VISIR spectra (except MCG--5--23--16; see below) have lower or equal fluxes than the \Spitzer spectra, which is expected for the higher spatial resolution. A similar result for VISIR photometry on a sample of type 1 and type 2 AGN has recently been reported by \citet{Hor09}. Together with the fact that the imaging-photometry is well consistent with the spectro-photometry, this confirms self-consistency of the extracted fluxes and supports consistency of our calibration approach with respect to other observations. 

\begin{figure*}
\centering
\includegraphics[width=1.0\textwidth]{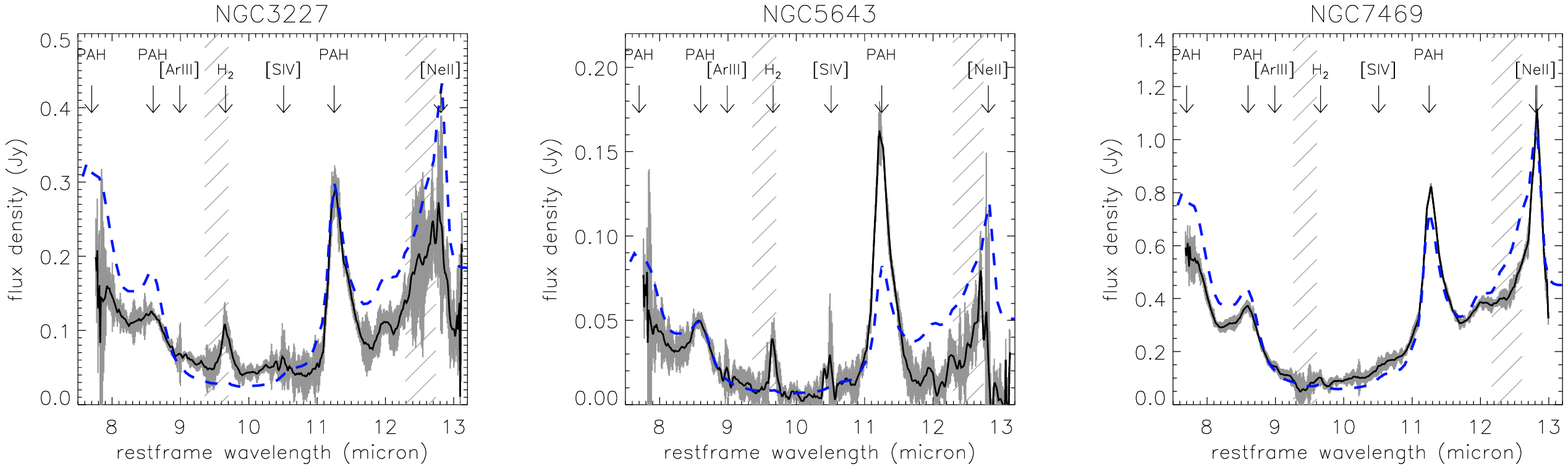}
\caption{\Spitzer IRS minus VISIR differential spectra of NGC~3227 (left), NGC~5643 (middle), and NGC~7469 (right). The differential spectra are compared to the scaled-down IRS spectrum of M82 (blue-dashed line) which is commonly used as a template for extragalactic star-formation.}\label{fig:pah}
\end{figure*}

In MCG--5--23--16 the VISIR spectrum is slightly above both the Spitzer spectrum and the VISIR photometry data point. Since the VISIR photometry matches the Spitzer fluxes we account this discrepancy to some unknown calibration error in spectroscopy (e.g. a slightly deviating spectral shape of the calibrator). In the extreme case around 12$\,\micron$, the difference is 50\,mJy or 7\% of the flux. In NGC~7469, the VISIR photometry is slightly but systematically lower the VISIR spectrum. The reason for this is that the extraction window used for the spectroscopic data includes some small degree of emission from the surrounding star-burst ring (see also Sect.~\ref{sec:sb} and \citet{Hor09}) while our optimal flux extraction for imaging isolated the nuclear point source.

As most ground-based mid-IR instruments, VISIR uses chopping and nodding to subtract the strong sky emission. However, this procedure does not only affect the background but also has an influence on the science data: If an emission region is extended on scales larger than the chop-throw, this emission is strongly reduced. The degree of suppression depends on the actual extension and spatial flux gradient. It is strongest for objects which are more or less uniformly emitting light. On the other hand, point-like objects or emission regions confined within the chop-throw are not affected. Thus, chopping can be considered a useful tool to reduce or suppress extended host galaxy emission. Such contribution from the host galaxy can be stellar light or emission from extended star-formation projected onto the nucleus, in particular if the host galaxy is strongly inclined. In conclusion, our spectra and photometry can be expected free of extended host galaxy emission, containing only emission from the point-like AGN and its immediate vicinity.

\subsection{PAH emission and star-formation activity around the AGN}\label{sec:sb}

Thirteen of the \Spitzer IRS spectra of our objects show more or less pronounced PAH emission features at $11.3\,\micron$, and in some cases the $7.7/8.6\,\micron$ PAH complex is also well visible. Since PAH emission is associated with star-formation, the presence of these features in the IRS data indicate that star-formation is occurring in the central $\sim3\arcsec$ of the AGN or, due to the lack of chopping, is projected onto the nucleus in the Spitzer data. Common to all of our high-spatial resolution VISIR spectra as compared to the \Spitzer IRS data is the lack, or at least strong suppression, of PAH line emission. Most objects do not display any $11.3\,\micron$ PAH emission, while in NGC~3227, NGC~5643, NGC~5995, NGC~7469, and NGC~7582 some minor PAH feature can be seen. For Mrk~509 the situation is inconclusive due to the low quality of the 12.4\,$\micron$ setting, and for NGC~4593 we do not have data in the corresponding spectral setting. In the cases with some little $11.3\,\micron$ PAH emission, it is difficult to judge if some marginally remaining feature at around 8.6\,$\micron$ is also present (e.g. by comparing to the Spitzer data), given that this feature is generally broader in wavelength and weaker. It has to be pointed out that the suppression of the PAH emission features is not an effect of different \textit{spectral} resolution but only caused by the different \textit{spatial} resolution and observing techniques. 

In Fig.~\ref{fig:pah}, we show examples of IRS minus VISIR differential spectra of objects with prominent PAH emission in the IRS spectra, NGC~3227, NGC~5643 and NGC~7469. For that, the VISIR data were downgraded to the same spectral resolution as IRS ($R\sim100$). The differential spectra are compared to the scaled-down version of the IRS spectrum of M82 (blue-dashed line), which is often used as an extragalactic star-formation template. The match for all three sources is reasonably good, especially for NGC~7469.
% The NGC~7469 difference spectrum is well reproduced by the M82 spectrum, while in NGC~3227 and NGC~5643 the M82-template is not exactly fitting the differential spectrum. However, the overall shape of spectrum and the lines contained are comparable, 
Thus, we conclude that the differential spectra from spatial regions between $\sim 0\farcs3 - 3\arcsec$ are predominantly showing star-formation emission in the continuum and lines.

The main differences of spectra taken with \Spitzer and VISIR are spatial resolution and observing technique. Since our spatial resolution is about a factor of 10 better than the IRS data, the emission regions producing the PAH features can simply be located at scales between 0$\farcs$3 and 3$\arcsec$. On the other hand, only in few cases individual star-forming regions are actually seen in the VISIR images, and images of these cases have recently been shown by \citet{Hor09}. The most notable example is NGC~7469 with its well-known star-burst ring at about 2$\arcsec$ from the nucleus. When integrating over a 0$\farcs$75 $\times$ 3$\arcsec$ aperture, we recover part of the IRS spectrum. The remaining ``missing flux'' probably originates from extended emission outside the VISIR aperture or from the host galaxy projected onto the nucleus. Since VISIR uses the chopping/nodding technique, any extended host galaxy emission at scales beyond the chop throw will be eliminated from the data or, at least, significantly reduced. As a result, the nuclear point source flux is free of contaminating emission from the host galaxy, even if part of the host emission falls onto the nucleus.

%\begin{figure}
%\centering
%\vspace{-0.5cm}
%\includegraphics[width=0.5\textwidth]{NGC7469_PAH2REF2_IMG.eps}
%\caption{VISIR 11.8\,$\micron$ image (PAH2ref2 filter) of the nuclear region of NGC~7469. The image was scaled with a power law index of 0.15 to decrease the contrast between starburst ring and nucleus. To obtain the final image, 3 individual observations from 2006 were co-added on the nucleus.}\label{fig:n7469img}
%\end{figure}

Still, most of the PAH emission in Spitzer is supposed to come from the vicinity of the AGN. Since we selected galaxies with only moderate inclination, projection effects should play a minor role. Consequently, we see a significant reduction of the PAH emission from scales of 1\,kpc down to $<$100\,pc. What is the reason for the suppression of the PAH emission features? PAH dust grains are prone to photo-destruction by high-energetic photons as emitted by an AGN or in its vicinity (e.g. from hot thermal plasma). Therefore it is quite plausible that they become less abundant at smaller distances from the AGN. On the other hand, a lack of PAH emission can also point to reduced or no star-formation activity. While it is not possible to distinguish between both effects only from the PAH emission, the reduction of PAH emission features from IRS to VISIR goes along with a reduction in continuum flux in our spectra. If the only reason for suppressed PAH emission were photo-destruction of the associated grains, we would expect that the continuum level remains roughly constant. Thus, we conclude that the dominating effect of suppressed PAH emission is an actual decrease in star-formation activity at smaller distances from the AGN. 

\subsection{The \boldmath$N$-band silicate feature in AGN at high spatial resolution}\label{sec:siobs}

Silicate absorption and emission features are the most evident spectral signatures in the $N$-band. One of the surprising discoveries of \Spitzer was based on observations of silicate emission features in type 1 AGN. The features have been much weaker than expected from early torus modeling \citep[e.g.][]{Pie93,Gra94,Efs95}. The same also holds for silicate features in absorption as seen in type 2 AGN, although very deep features are sometimes seen in ULIRGs or type 2 AGN where the host galaxy presumably contributes by a large fraction to the obscuration \citep[e.g.][]{Lev07,Pol08,Mar09}. Since the \Spitzer data has a spatial resolution of several arcseconds in the $N$-band, it was not clear initially to what degree the weakness of the silicate features was a resolution effect. Recent ground-based mid-IR spectroscopy of single objects at high spatial resolution, however, suggested that even at resolution better than 1$\asec$ silicate absorption and emission features were moderate \citep[e.g.][]{Roc07,Mas09}. Here we present a much larger sample of type 1 and type 2 AGN observed at sub-arcsecond resolution so that the silicate feature characteristics can be studied more systematically (see Sect.~\ref{sec:si}).

\begin{figure}
\centering
\includegraphics[width=0.5\textwidth]{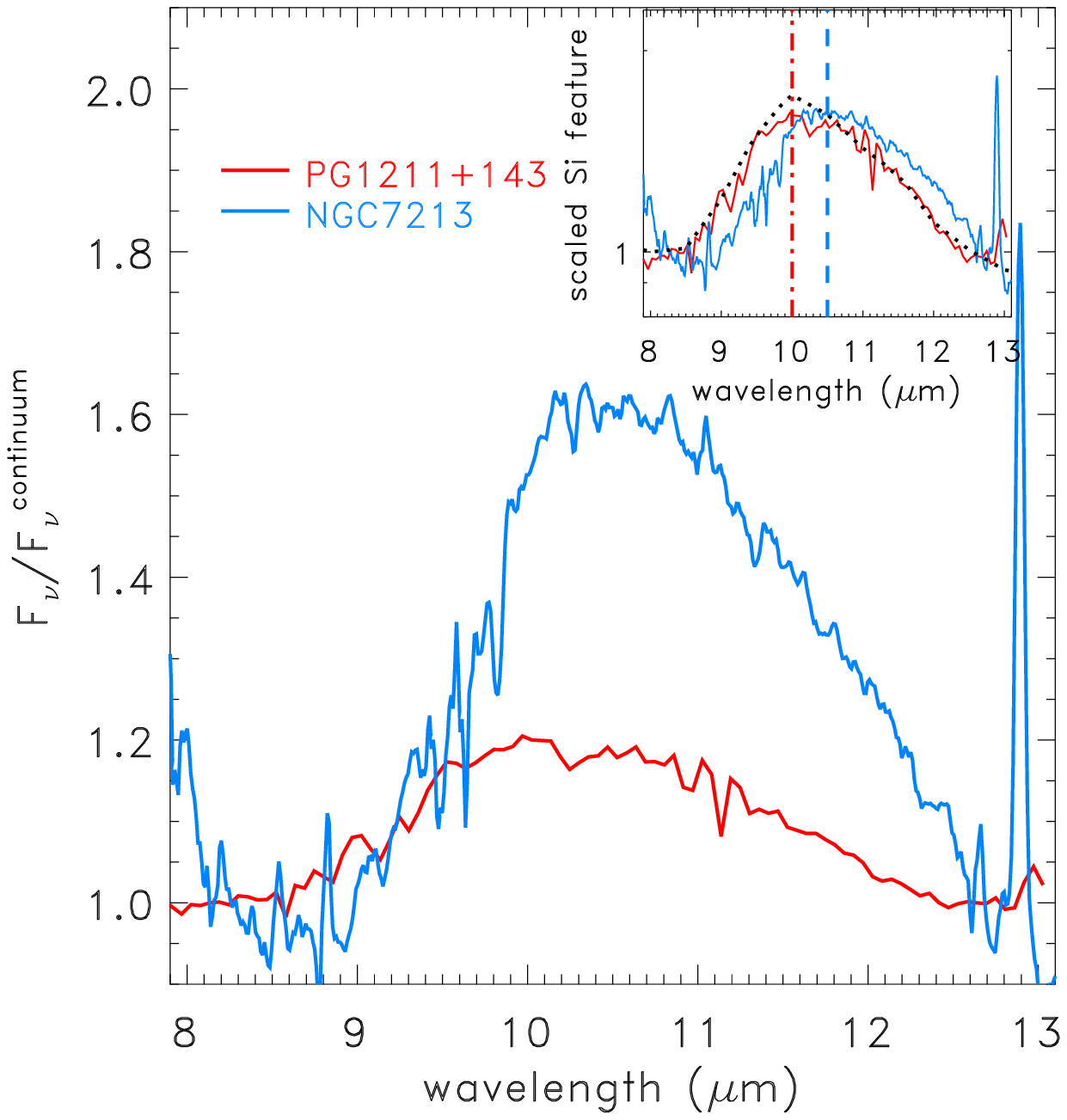}
\caption{Continuum-normalized silicate emission features of PG1211+143 (red) and NGC~7213 (blue). A linear continuum has been fitted to both objects based on their 8.3 and 12.7$\,\micron$ fluxes and the observed spectrum was divided by this continuum fit. The small inset shows both features in log plot normalized for their different feature strength. As a comparison we overplot the extinction coefficient of standard ISM dust with Ossenkopf silicates (black-dotted line). The blue-dashed and red-dotted-dashed lines denote the central wavelengths of NGC~7213 and PG1211+143 respectively.}\label{fig:sifeat}
\end{figure}

Fig.~\ref{fig:VISIR_type1} illustrates that most type 1 AGN exhibit a rather weak emission feature, if we can detect any at all. A small bump in the continuum can be seen somewhere between 9 and 12\,$\micron$, e.g. in NGC~3783 and MCG--6--30--15. The other features are only revealed when plotting $\log \nFn$ or fitting the continuum (see Sect.~\ref{sec:si}) and some seem to be rather \textit{absorption} than emission features (e.g. NGC~3227). The strongest silicate emission features is displayed by the weak or LINER AGN NGC~7213 (Fig.~\ref{fig:VISIR_type1_L}). Based on the spatial resolution of our data of $<$100\,pc in all objects except MARK~509, we conclude that the weakness of the silicate emission features in type 1 AGN is not a spatial resolution effect but an intrinsic property of the features and possibly of the emission region.

The type 2 AGN show a variety of feature characteristics (see Fig.~\ref{fig:VISIR_type2}). While most type 2s have moderate silicate features in absorption, the NGC~4507 $N$-band spectrum resembles a featureless type 1 spectrum. In NGC~2110, we even see the silicate feature in emission. It is not surprising seeing such different characteristics in type 2s since we expect various torus inclinations realized in these objects, ranging from moderate to very high obscuration (``edge-gracing'' to edge-on geometries). This is illustrated by the variety of Hydrogen column densities $\nh$ observed in X-rays for these AGN (see Table~\ref{tab:scales2}). In fact, NGC~2110 has the lowest $\nh$ within the type 2 sub-sample and infrared broad emission lines have been reported \citep{Ver06} \citep[see][for possible scenarios in NGC~2110]{Mas09}.

One of the mysteries of the silicate feature in AGN is an actual or apparent shift of the emission feature towards longer wavelengths. Absorption features in type 2 AGN are centered very close to $9.7\,\micron$, just as expected from opacity curves of the Galactic ISM dust \citep{Chi06}. Clumpy torus models based on ISM dust well reproduce the overall IR dust re-emission and the silicate absorption features in detail \citep[e.g.][]{Hon06,Hon07a,Pol08,Sch08}. For type 1 AGN, the situation is less clear. Early reports of the silicate emission feature in a number of quasars discussed that the central wavelength might be shifted towards longer wavelengths \citep{Hao05,Sie05,Stu05}. Recently, \citet{Nik09} suggested that radiative transfer effects might cause the silicate emission feature in AGN to shift towards longer wavelengths. In their analysis, they require a certain number of clouds to intervene the line-of-sight of other clouds to cause some absorption tip at the center of the silicate feature. To occur at the right wavelength, they suggest using \citet{Oss92} silicates with peak wavelength at 10.0\,$\micron$ instead of standard ISM. However, they note that any slightly larger number of clouds or different opacity would cause a noticeable absorption dip to be present in the center of the silicate emission feature, as shown in the models of \citet{Nen08b} and \citet{Hon10}. Based on our own modeling, we would expect that if such radiative transfer effects are the \textit{only} reason for a shift or the silicate emission feature, then we would expect that analyzing a sample of objects would reveal both shifted features and those showing noticeable absorption dips within the emission feature. However, all of the type 1 AGN observed here and in addition NGC~2110 as a type 2 AGN with a silicate emission feature do \textit{not} show such an absorption dip within the emission feature. This is consistent with a detailed study of the silicate emission features of 23 PG quasars observed with \Spitzer \citep{Swe08}. The problem was also mentioned by \citet{Mas09} when attempting to model the silicate emission feature in NGC~2110. In addition, we note that our modeling of silicate features with Ossenkopf et al. silicate opacities show both emission \textit{and} absorption features centered at 10.0\,$\micron$, in agreement with \citet{Nen08b}. Since, however, the observed absorption feature central wavelength is at 9.7\,$\micron$ \citep[see also][]{Mas06,Roc07,Hon07a}, type 2 AGN would favor different opacity curves. 

In Fig.~\ref{fig:sifeat} we show a comparison between the continuum-normalized $N$-band \Spitzer spectrum of PG1211+143 analyzed by \citet{Nik09} and our VISIR spectrum of NGC~7213. To estimate the continuum, we made a linear fit to the 8.3 and 12.7\,$\micron$ fluxes in $F_\nu$. The observed spectra are then divided by the continuum fit. Using this kind of fit (which is to some extent similar to the method suggested by \citet{Sir08} and used by \citet{Nik09}), we see that the peak emission of PG1211+143 is located indeed at 10.0\,$\micron$, while it is shifted to $\sim$10.5\,$\micron$ in NGC~7213. We compare these features to the extinction curve of standard ISM dust with Ossenkopf silicates and find that the PG1211+143 silicate emission profile is well matched while a shift is evident in NGC~7213. Note that if we used the \Spitzer IRS spectrum of NGC~7213 instead of the VISIR spectrum, the result would be more or less the same.

In summary, we would conclude that simple radiative transfer effects due to absorption within the torus \textit{alone}, either clumpy or smooth, are not capable of explaining the details of the silicate features and their characteristics in type 1 and type 2 AGN. There may be other effects contributing to the actual shape of the feature. One possibility is that there is some (radial) change of the grain size and/or dust composition based on the fact that large grains, preferably graphite grains, have a much higher sublimation temperature than smaller grains, in particular silicates. Then, hotter regions might be dominated by graphite dust and larger grains while cooler regions have an ISM-like size and graphite-silicate mix, which may go along with a temperature-dependent central wavelength of the silicate feature \citep[e.g. as illustrated in][Fig.~9.7]{Kru08}. Consequently, exact details on absorption and emission within the silicate feature are more complex than covered by recent torus models, although the essence of the features (i.e. feature strength) are well captured by these models.

\section{AGN continuum and dust emission properties in the mid-IR at \boldmath$<$100\,pc}\label{sec:plots}

As shown in the previous section, the high-spatial resolution of VISIR allows us to isolate the nuclear mid-IR emission in AGN without too much disturbance from host-galactic sources. Thus, analysis of this data will reveal characteristics of the circumnuclear region around the AGN. In the following, we will discuss several properties of the mid-IR continuum emission and the silicate features. Narrow forbidden atomic emission lines which are also present in the spectra, like [\ion{Ar}{iii}], [\ion{S}{iv}], and [\ion{Ne}{ii}] have been already studied in part in \citet{Hon08}, and the whole sample will be dealt with separately in an upcoming paper.

\begin{figure}
\centering
%\vspace{-0.5cm}
\includegraphics[width=0.5\textwidth]{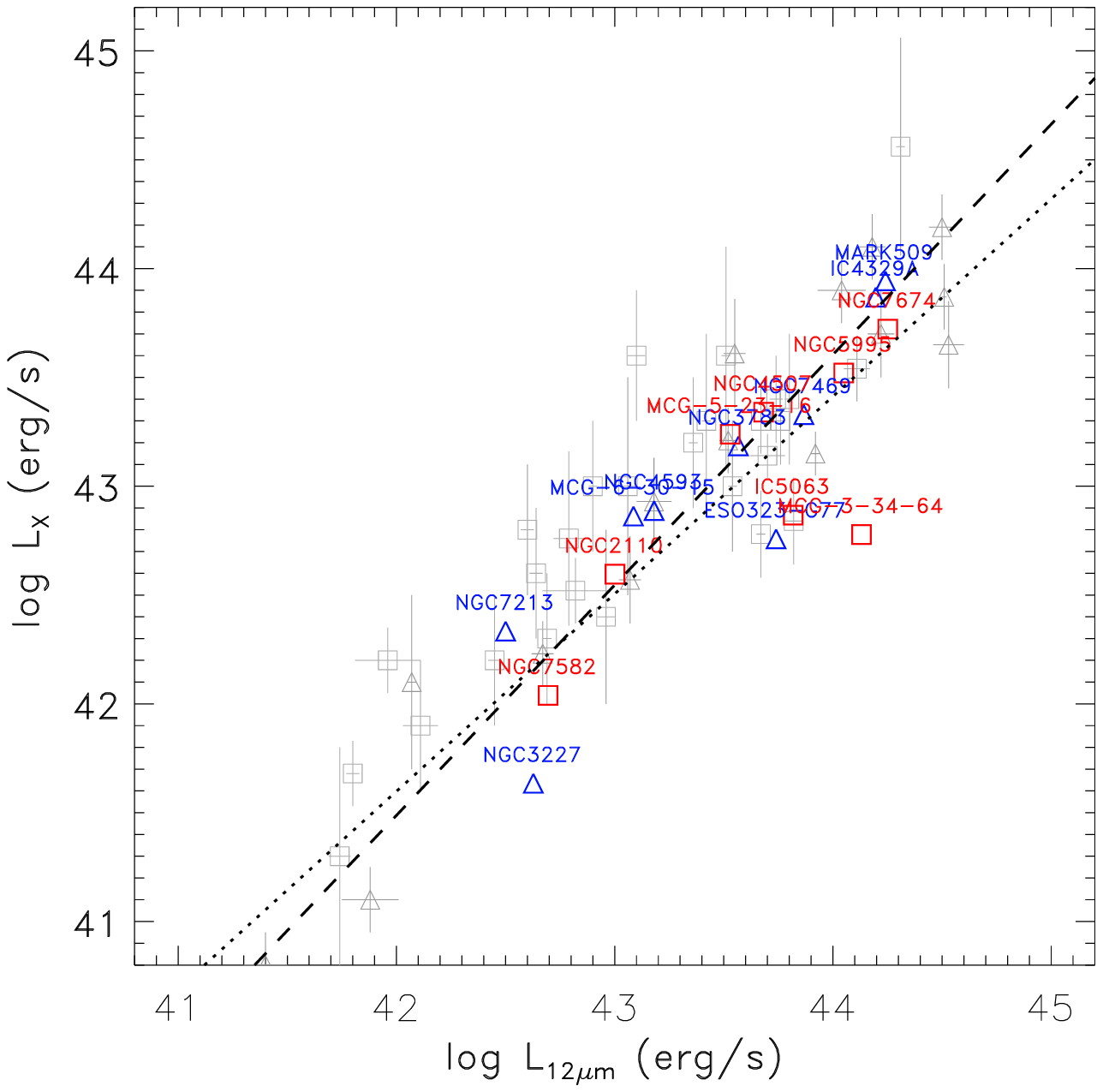}
\caption{Absorption-corrected X-ray luminosities $\LX$ of our sample of AGN plotted against the $12\,\micron$ luminosity $\Lm$ derived from the VISIR spectra, excluding Compton-thick objects (see Tables~\ref{tab:scales} \& \ref{tab:scales2}). The blue triangles are type 1 objects while the red squares are type 2 AGN. The dotted line is the best-fit correlation for our whole sample, while the dashed line represents a fit to the sample excluding the outliers (see Sect.~\ref{sec:lmircont}) The gray symbols show the full sample of the latest version of the $\LmLx$-correlation as presented in \citet{Gan09}.}\label{fig:LmLx}
\end{figure}

\subsection{Mid-IR continuum emission and its relation to AGN luminosity tracers}\label{sec:lmircont}

The mid-IR emission is believed to originate from dust reprocessing of the central accretion disk radiation. In particular, optical/UV photons hit the dust in the torus and heat up the grains. The absorbed energy is then re-emitted in the IR and forms the general IR bump seen in multi-wavelength SEDs \citep[e.g.][]{Elv94}. This redistribution of energy should produce a fairly significant correlation between the X-ray (accretion disk tracer) and mid-IR (dust emission tracer) luminosity in AGN. Such a $\LmLx$-correlation has been found by several authors using both space-based and ground-based observations \citep[e.g.][]{Kra01,Lut04,Hor08,Gan09}. One remarkable property of this correlation is the apparent isotropy among the AGN: Within the observational scatter of the latest version of the correlation of about a factor of 2 to 3 \citep{Gan09}, there seems to be no difference between type 1 and type 2 AGN.

\begin{figure*}
%\sidecaption
\centering
\includegraphics[width=0.9\textwidth]{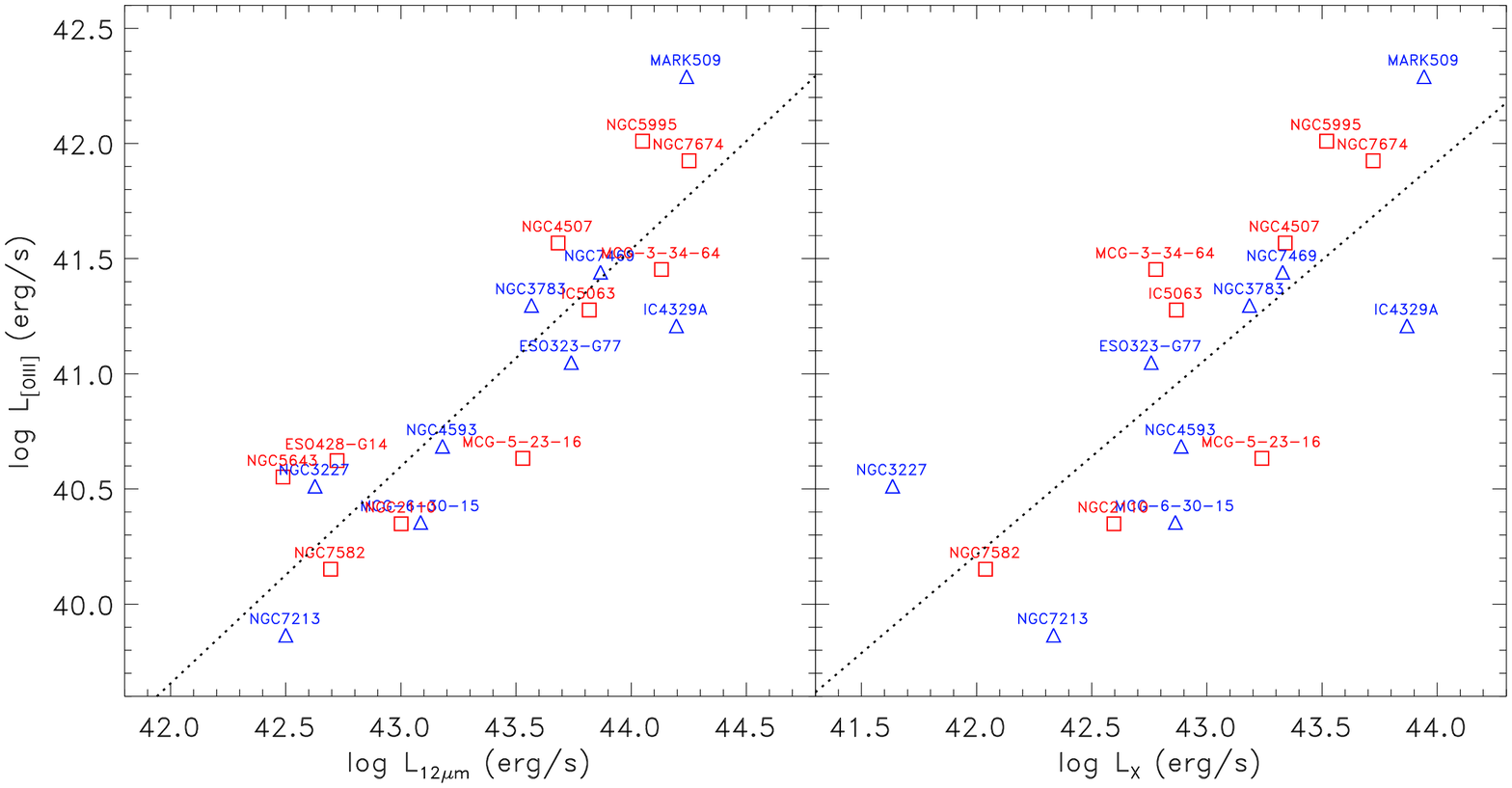}
\caption{[\ion{O}{iii}] luminosities of our sample compared to the $12\,\micron$ mid-IR (left) and $2-10$\,keV X-ray luminosities. The blue triangles are type 1 AGN, the red squares denote type 2 objects. The dotted lines are best-fit correlations for the whole sample (left: $\Loiii \propto \Lm^{0.96\pm0.12}$; right: $\Loiii \propto \LX^{0.85\pm0.18}$).}\label{fig:loiii}
\end{figure*}

In Fig.~\ref{fig:LmLx} we show the absorption-corrected X-ray luminosities $\LX$ of our sample of AGN plotted against the $12\,\micron$ luminosity $\Lm$ derived from the VISIR spectra (see Tables~\ref{tab:scales} \& \ref{tab:scales2}). Compton-thick objects are not shown in the plot. The errors of the mid-IR luminosities are $<$0.1 dex (typical photometric accuracy is much better than 15\%). The X-ray luminosities are taken from single-epoch data, so that we consider typical uncertainties in $\LX$ due to intrinsic variability of $\sim$0.5\,dex (see error of the \citet{Gan09} sample overplotted as gray symbols in Fig.~\ref{fig:LmLx}). Within the scatter of our observations, there is no apparent difference between type 1 and type 2 AGN. This is consistent with the larger samples analyzed by \citet{Hor08} and \citet{Gan09}; the latter sample is shown for comparison as gray symbols in Fig.~\ref{fig:LmLx}. Note that some objects in the original Gandhi et al. sample are duplicated in our sample. However, different X-ray luminosities have been used and the mid-IR fluxes were extracted from our VISIR spectra. Using our much smaller sample leads to a nominal correlation $\Lm \propto {\Lx}^{1.06\pm0.15}$ (Spearman rank $\rho_\mathrm{Spearman}=0.82$, null-hypothesis probability $4.9\times10^{-5}$) which is consistent with the findings in \citet{Gan09}. There seems to be a small offset of our sample towards lower mid-IR luminosities with respect to the Gandhi et al. study. This can be explained to a small degree by the fact that Gandhi et al. used $12.3\,\micron$ as the reference wavelength where fluxes are usually slighly higher than at $12.0\,\micron$ as used here. Most of the difference, however, comes from the method used to extract the fluxes: Our photometry was measured using the optimal extraction while in \citet{Gan09} aperture photometry was used which potentially includes some off-nuclear emission.

The non-detection of a difference between type 1 and type 2 AGN within the error of observations might indicate a relatively large degree of isotropy of the AGN radiation in both the mid-IR and the X-ray. However, it has to be noted that the $\LmLx$-correlation omits all Compton-thick objects. Unless one has a good idea of the intrinsic X-ray luminosity of these highly obscured objects \citep[e.g. from line emission at optically thin wavelengths as in][]{Hon08}, the observed isotropy must be considered as a lower limit. One such possibility is the commonly used [\ion{O}{iii}]($\lambda5007\,\mathrm{\AA}$) luminosity, although some degree of anisotropy has been reported \citep[e.g.][]{Net06}. In Fig.~\ref{fig:loiii} we compare the [\ion{O}{iii}] luminosities of our sample to the $12\,\micron$ mid-IR and $2-10$\,keV X-ray luminosities, respectively. Both properties correlate well with $\Loiii$, but the $\Lm-\Loiii$-relation is stronger by eye and by a statistical analysis. For our sample we find $\log \Loiii = (0.11\pm5.3) + (0.94\pm0.12)\times\log\Lm$ (Spearman rank $\rho=0.87$, null-hypothesis probability $1.3\times10^{-6}$) and $\log \Loiii = (4.38\pm7.8) + (0.85\pm0.18)\times\log\LX$ (Spearman rank $\rho=0.78$, null-hypothesis probability $1.9\times10^{-4}$). Note that for the correlation with $\Lm$, there are also Compton-thick objects included (NGC~5643 and ESO~428--G14). Thus, if the [\ion{O}{iii}] radiation is considered as being a good isotropy tracer, then $\Lm$ may be considered as being emitted quite isotropically as well -- at least more isotropically than X-ray radiation. This might also be interpreted as a sign that the mid-IR optical depth of any obscuring medium (e.g. as traced by the X-ray hydrogen column density $\nh$) must be more transparent in the mid-IR, or that the mid-IR emission is emitted within or outside the X-ray-opaque medium.

These findings seem to have a natural interpretation in the unification scheme: The dusty torus is both the X-ray-obscuring and the mid-IR-emitting medium. Recent studies have shown that the observed mid-IR isotropy (i.e. small dispersion between type 1 and type 2 AGN in the $\LmLx$-correlation) can be explained within the framework of a clumpy torus \citep[e.g.][]{Hon06,Hor08,Nen08b,Gan09,Lev09}. There is, however, an interesting point to consider: While IC~5063, ESO~323--G77, and MCG--3--34--64 are outliers in the $\LmLx$-correlation \citep[see Fig.~\ref{fig:LmLx} and also][]{Hor08,Gan09}, they are very close to the $\Lm-\Loiii$-fit (see Fig.~\ref{fig:loiii}). This could be caused by absorption since all three objects have moderate $\nh$ ($\log\nh \sim 23$). On the other hand, all recent $\LmLx$-studies use $\nh$-corrected $\Lx$ so that it is difficult to imagine that the intrinsic $\Lx$ are still underestimated by a factor of $\sim$3$-$5. Another possibility would be that these objects are not ``underluminous'' in the X-rays but ``overluminous'' in the mid-IR, i.e. there is additional mid-IR emission other than the dust torus re-emission. This overhead emission is probably triggered by the AGN, since our spectra do not include any major star-burst component (see Sect.~\ref{sec:sb}). Instead the additional mid-IR emission may originate from dust in the narrow-line region (NLR). The prime example for such a case is NGC~1068. \citet{Mas06} point out that the flux in the central 1$\farcs$2 of this AGN ($\sim$80\,pc, which is a typical resolution element of our sample) contains only 30\% contribution from the nuclear point source (presumably the torus) while 70\% originate in extended emission, mostly from the NLR\,\footnote{We note that the NGC~1068 data used in Figs~/ref{fig:specindex} to \ref{fig:model} are point source values from a 0$\farcs$4 extraction window.}. This corresponds to a mid-IR ``overluminosity'' of 0.3$-$0.4\,dex in the $\LmLx$-correlation, which is about the offset of the three sources. If we exclude these sources from the $\LmLx$-correlation analysis, we obtain a tight relation of $\Lm \propto {\Lx}^{0.94\pm0.10}$ (Spearman rank $\rho_\mathrm{Spearman}=0.96$, null-hypothesis probability $2.5\times10^{-8}$), which is consistent within errors with \citet{Gan09}.

\begin{figure}
%\sidecaption
\centering
\includegraphics[width=0.5\textwidth]{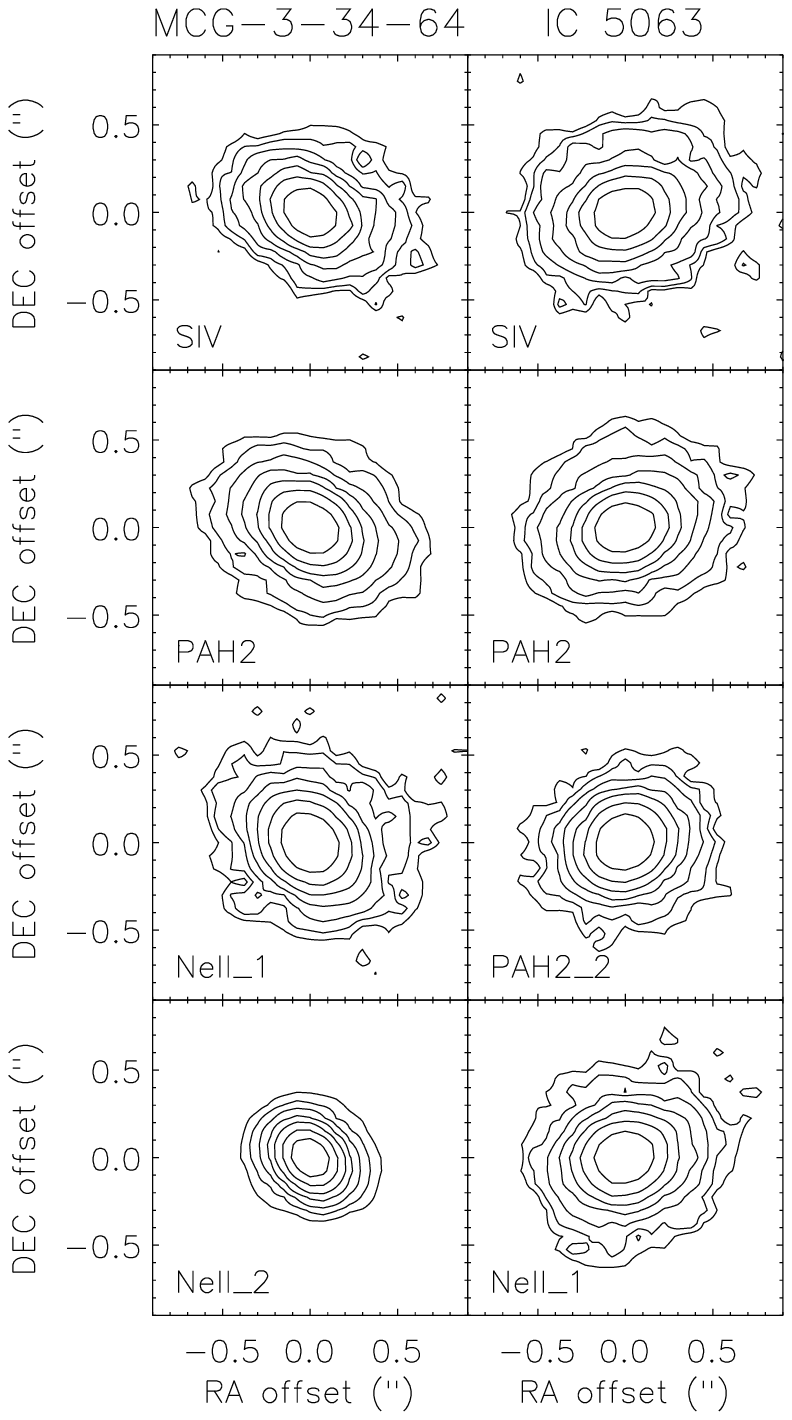}
\caption{VISIR mid-IR contour images of MCG--3--34--64 (left column) and IC~5063 (right column). The different panels in each column represent images in different filters (SIV = forbidden [\ion{S}{iv}](10.51\,$\micron$) line, PAH2 = 11.3\,$\micron$ PAH line, PAH2\_2 = 11.88\,$\micron$ continuum, NeII\_1 = 12.27\,$\micron$ continuum, NeII\_2 = 13.04\,$\micron$ continuum). Contour levels are scaled logarithmically from the peak in steps of 0.2dex down to the 4\% peak flux level. Both objects show clear elongations, corresponding geometric properties based on Gaussian fits as listed in Table~\ref{tab:extend}.}\label{fig:extend}
\end{figure}

An interesting aspect about this suggestion is that mid-IR VISIR imaging of the outliers MCG--3--34--64 and IC~5063 actually shows extended emission. \citet{Hor09} mention some ``slight elongation'' in their SIV, PAH2, and NeII\_1 filter images of MCG--3--34--64, but attribute it to instrumental effects. We reanalyzed their images of MCG--3--34--64 and IC~5063, together with additional archival VISIR images (see Fig.~\ref{fig:extend}), by applying a 2D Gaussian fit to both science target and calibrator. The resulting FWHM and position angles are presented in Table~\ref{tab:extend}. MCG--3--34--64 shows at least 38--54\% and IC~5063 shows at least 14--45\% extension compared to the maximum elongation of the respective calibrators. All calibrators appear more or less round with axis ratios close to unity. On the other hand, the average axis ratios are 1.35 for MCG--3--34--64 and 1.26 for IC~5063. This goes in line with very similar position angles in each filter while the orientation of the axis in the standard star images is more random (average PA 54$^\circ$ for MCG--3--34--64 and 111$^\circ$ for IC~5063). In IC~5063, this is consistent with the orientation of the extended [\ion{O}{iii}] emission as observed with HST \citep[PA~115$^\circ$;][]{Stt03}. Despite no high-spatial resolution [\ion{O}{iii}] image is available for MCG--3--34--64, we can compare our results to radio observations. \citet{Sch01} report linear extension of the nuclear radio emission (inner $\sim$100\,pc) at 8.46\,GHz towards PA 39$^\circ$ which is only about 13$^\circ$ off from the mean elongation of the mid-IR emission. If we assume that the linear radio emission traces some jet-like emission and that the jet-like emission is approximately in the same direction as the NLR, we can argue that the mid-IR emission in MCG--3--34--64 is also extended in the NLR direction.

Actually, these 2 objects are the only ones in our sample where we see such a significant extension and elongation (see data in Tables~\ref{tab:obs} \& \ref{tab:obs2}; point-like images for some of our objects are shown in \citet{Hor09}). Moreover, some members of our group attempt VLTI/MIDI mid-IR interferometry on both objects in the past. Although total fluxes were easily recorded, no fringes have been found. This is consistent with a very low 12\,$\micron$ visibility (0.2$-$0.3 at about 60\,m baselines) in these objects, meaning that 70--80\% of the nuclear flux comes from extended emission. Thus, based on (1) the absence of PAH emission lines in our VISIR spectra, (2) the resolved emission regions observed in the mid-IR images of MCG--3--34--64 and IC~5063, (3) the significant elongation, (4) consistent position angles of the mid-IR extended emission region and the [\ion{O}{iii}] emisison in IC~5063 and the linear radio emission in MCG--3--34--64, and (5) the low visibilities in mid-IR interferometry, we suggest a scenario where theses two objects have very mid-IR-bright NLRs -- similar to NGC~1068 -- which causes them to deviate from the $\LmLx$-correlation. Whether or not this may also be an explanation for the apparent ``overluminosity'' of AGN at the low-luminosity end of the $\LmLx$-correlation \citep{Hor08,Gan09} has to be addressed by detailed mid-IR studies.

\begin{table}
\caption{Gaussian fit geometric properties of MCG--3--34--64 and IC~5063 base on VISIR images}\label{tab:extend}
\centering
\begin{tabular}{l c c c c}
\hline\hline\
Filter   & \multicolumn{2}{c}{object properties} & \multicolumn{2}{c}{calibrator properties} \\
         & size         & PA      & size           & PA \\ \hline
\multicolumn{5}{c}{\bf MCG--3--34--64} \\
SIV      & $0\farcs488\times0\farcs342$ & 59$^\circ$ & $0\farcs316\times0\farcs298$ & 95$^\circ$ \\
PAH2     & $0\farcs514\times0\farcs367$ & 57$^\circ$ & $0\farcs355\times0\farcs320$ & 93$^\circ$ \\
NeIIref1 & $0\farcs505\times0\farcs400$ & 45$^\circ$ & $0\farcs365\times0\farcs328$ & 82$^\circ$ \\
NeIIref2 & $0\farcs562\times0\farcs435$ & 46$^\circ$ & $0\farcs391\times0\farcs384$ & 129$^\circ$ \\ \hline
\multicolumn{5}{c}{\bf IC~5063} \\
SIV      & $0\farcs551\times0\farcs417$ &110$^\circ$ & $0\farcs380\times0\farcs373$ & 175$^\circ$ \\
PAH2     & $0\farcs508\times0\farcs382$ &109$^\circ$ & $0\farcs379\times0\farcs368$ & 69$^\circ$ \\
PAH2ref2 & $0\farcs452\times0\farcs394$ &117$^\circ$ & $0\farcs397\times0\farcs385$ & 36$^\circ$ \\
NeIIref1 & $0\farcs485\times0\farcs394$ &107$^\circ$ & $0\farcs384\times0\farcs380$ & 160$^\circ$ \\ \hline
\end{tabular}
\end{table}

\subsection{The mid-IR spectral index}\label{sec:specind}

\begin{figure*}
%\sidecaption
\centering
\includegraphics[width=0.9\textwidth]{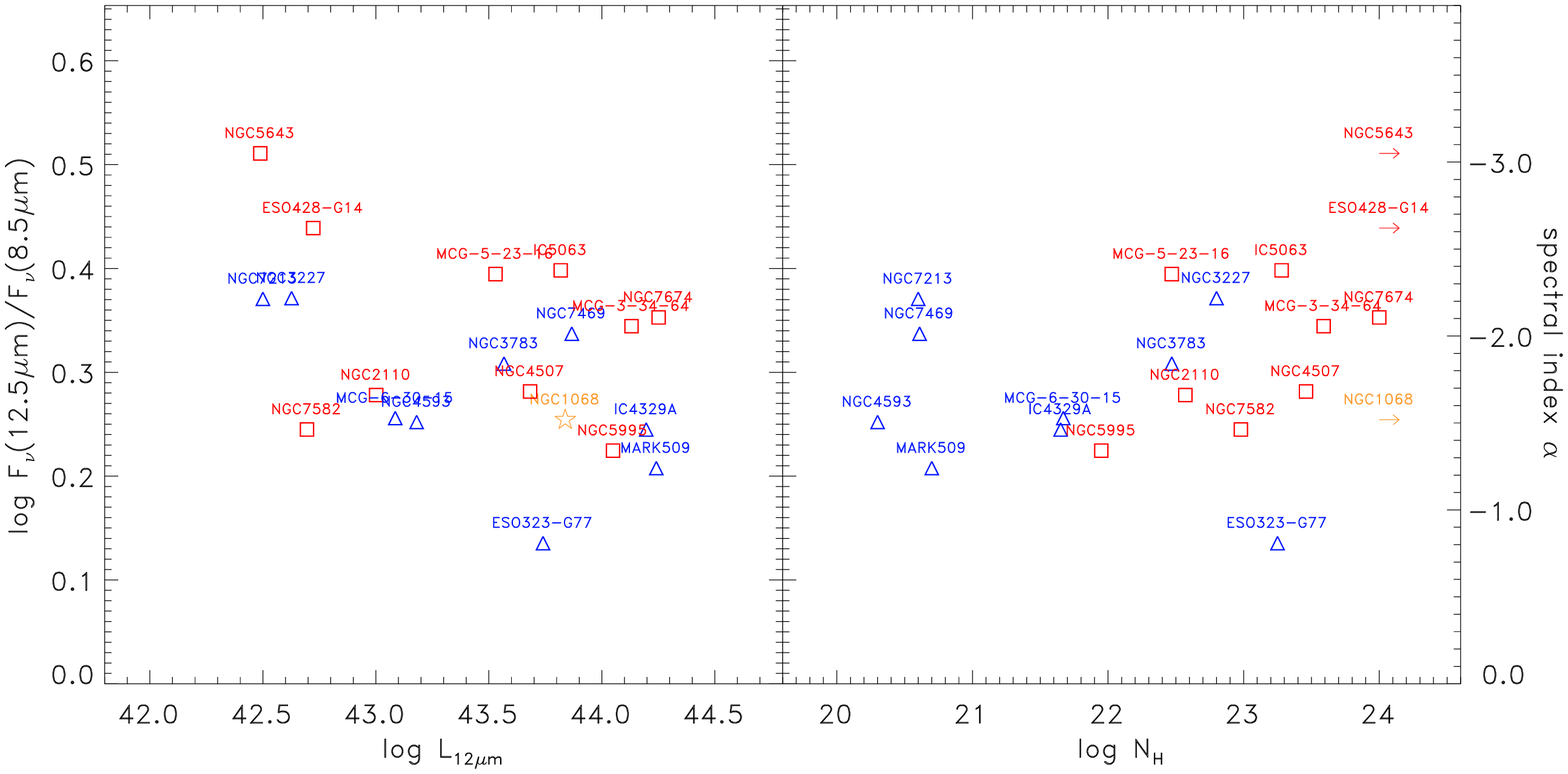}
\caption{Mid-IR spectral slope and spectral index for each object in our sample, plotted against the observed mid-IR luminosity $\Lm$ and the X-ray Hydrogen column density $\nh$. Type 1 AGN are shown as blue triangles, type 2 objects are marked by red squares. For reference, NGC~1068 is shown as an orange star, based on the mid-IR spectrum from \citet{Mas06}.}\label{fig:specindex}
\end{figure*}

One fundamental observational property of the mid-IR emission is its spectral slope. Since a sizable part of the $N$-band is affected by the silicate feature, the continuum slope has to be recovered from wavelengths outside the feature, which is centered around 10\,$\micron$. Here we assume that fluxes at 8.5\,$\micron$ and 12.5\,$\micron$ are mostly unaffected by the silicate feature (which we tested by inspecting ISM dust extinction curves and \Spitzer IRS data covering a larger wavelength range). 

In Fig.~\ref{fig:specindex} we show the spectral slopes (= flux ratios $F_\nu(12.5\,\micron)/F_\nu(8.5\,\micron)$) and spectral indices $F_\nu \propto \nu^\alpha$ (based on the flux ratios) for all of our objects. In the left panel, we plot the spectral slope $\log F_\nu(12.5\,\micron)/F_\nu(8.5\,\micron)$ against the observed mid-IR luminosity. The distribution is rather wide, so that a clear correlation cannot be seen. However, there might be a slight trend that the reddest objects in both the type 1 and type 2 sub-sample are on the lower luminosity end of the plot. What can clearly be seen is that type 1 and type 2 AGN do not show a huge difference in the spectral index $\alpha$, i.e. type 1s can be as red as type 2 AGN, although a marginal difference may be detected. The nominal mean spectral indices are $\alpha_\mathrm{AGN1}=-1.65\pm0.44$ for type 1 AGN and $\alpha_\mathrm{AGN2}=-2.07\pm0.54$ for type 2 AGN. This almost similarity in spectral index becomes even more evident when comparing the Hydrogen column density $\nh$ with the spectral index. There is at best a marginal trend of the spectral index with obscuration towards the AGN -- the most obscured objects appear to be redder than the mean in this sample. On the other hand, NGC~1068 as a Compton-thick object has $\alpha=-1.5$ in the mid-IR, so that it is placed in the bulk of the other objects (see star in Fig.~\ref{fig:specindex}, left). The spectral index for each object has been tabulated in Table~\ref{tab:specprop}.

\begin{table}
\caption{Properties of the VISIR mid-IR spectra. For each object we list the 8.5$-$12.5\,$\micron$ spectral index $\alpha$, the silicate feature strength $F_\mathrm{Si}/F_\mathrm{c}$, and the corresponding optical depth in the silicate feature $\tau_\mathrm{silicate}$ (see Sects.~\ref{sec:specind} \& \ref{sec:si} for details)}\label{tab:specprop}
\centering
\begin{tabular}{l c c c}
\hline\hline
Object & $\alpha$ & $\log F_\mathrm{Si}/F_\mathrm{c}$ & $\tau_\mathrm{silicate}$ \\ \hline
\textit{NGC~1068} & \textit{--1.52} & \textit{--0.16} & \textit{0.38} \\
      NGC~2110 & $-1.66$ & $ 0.01$ & $-0.03$ \\
   ESO~428--G14 & $-2.62$ & $-0.25$ & $ 0.57$ \\
  MCG--5--23--16 & $-2.36$ & $-0.14$ & $ 0.32$ \\
      NGC~3227 & $-2.22$ & $-0.06$ & $ 0.13$ \\
      NGC~3783 & $-1.84$ & $-0.03$ & $ 0.06$ \\
      NGC~4507 & $-1.68$ & $ 0.02$ & $-0.05$ \\
      NGC~4593 & $-1.51$ & $ 0.06$ & $-0.15$ \\
   ESO~323--G77 & $-0.81$ & $ 0.02$ & $-0.05$ \\
  MCG--3--34--64 & $-2.06$ & $-0.09$ & $ 0.20$ \\
  MCG--6--30--15 & $-1.53$ & $ 0.03$ & $-0.08$ \\
      IC~4329A & $-1.46$ & $-0.01$ & $ 0.02$ \\
      NGC~5643 & $-3.05$ & $-0.20$ & $ 0.45$ \\
      NGC~5995 & $-1.34$ & $-0.05$ & $ 0.13$ \\
      MARK~509 & $-1.24$ & $ 0.11$ & $-0.25$ \\
       IC~5063 & $-2.38$ & $-0.12$ & $ 0.28$ \\
      NGC~7213 & $-2.21$ & $ 0.10$ & $-0.23$ \\
      NGC~7582 & $-1.46$ & $-0.45$ & $ 1.03$ \\
      NGC~7469 & $-2.01$ & $ 0.01$ & $-0.02$ \\
      NGC~7674 & $-2.11$ & $-0.08$ & $ 0.19$ \\ \hline
\end{tabular}
\end{table}

\begin{figure*}
%\sidecaption
\centering
\includegraphics[width=1.0\textwidth]{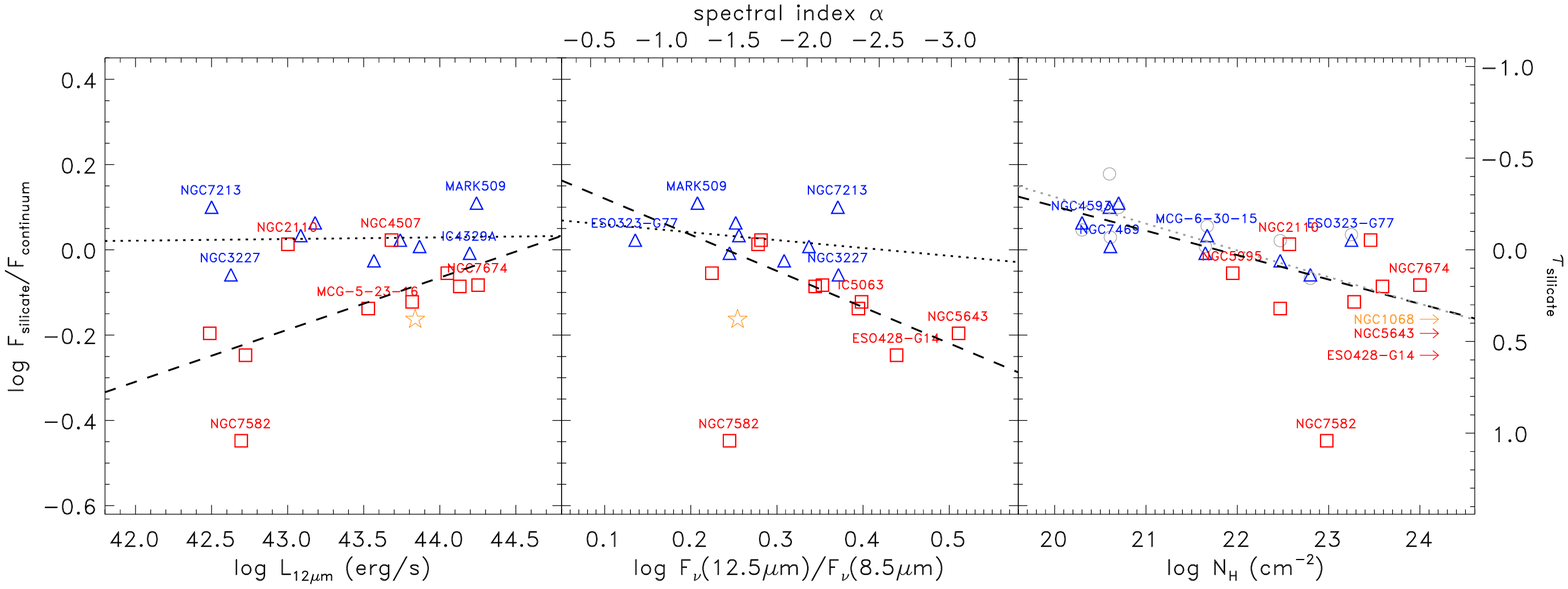}
\caption{Silicate feature analysis of our AGN sample observed with VISIR. Type 1 AGN are shown as blue triangles, type 2 AGN are marked by red squares. Some objects have been identified by name. For reference, we include NGC~1068 with mid-IR properties based on the Gemini spectrum by \citet{Mas06}. \textit{Left:} Dependence of the silicate feature strength $F_\mathrm{Si}/F_\mathrm{c}$ (see text for details) on the mid-IR luminosity $\Lm$. The dashed line shows the nominal relation for type 2 AGN, the dotted line is the corresponding relation for type 1s. \textit{Middle}: $F_\mathrm{Si}/F_\mathrm{c}$ plotted against the spectral slope and spectral index of the mid-IR continuum emission. The dashed line shows the nominal relation for type 2 AGN, the dotted line is the corresponding relation for type 1s. \textit{Right:} $F_\mathrm{Si}/F_\mathrm{c}$ and observed optical depth in the silicate feature $\tau_\mathrm{silicate}$ shown as a function of the Hydrogen column density $\nh$. The dashed line represents the correlation $F_\mathrm{Si}/F_\mathrm{c} \propto 0.12\,\times\,\log\nh$. Gray circles are type 1 data points if the central wavelength of the silicate emission feature is 10.3\,$\micron$ instead of 9.8\,$\micron$. The corresponding correlation $F_\mathrm{Si}/F_\mathrm{c} \propto 0.14\,\times\,\log\nh$ is shown as dotted line.}\label{fig:siprop}
\end{figure*}

The fact that type 1 and type 2 AGN have very similar spectral indices might seem quite surprising. The first-order assumption based on the torus picture would be that type 1 AGN are, on average, bluer than type 2 AGN, because most of the IR emission we see in type 1s comes from hot dust in the inner part of the torus. On the other hand, as shown in \citet[][e.g. Sects. 3.2, 3.4 \& 3.5.1, and Figs. 6, 7, 9, 10, 11 \& 12]{Hon10}, the way the dust is distributed around the AGN can have a much stronger effect on the mid-IR properties than inclination effects -- in particular if the torus is clumpy so that transitions in obscuration properties from type 1 to type 2 are smooth. We demonstrated that if the radial dust distribution is steep ($\eta_r\propto r^a$ with $a=-1.5\ldots-2.0$), i.e. most of the dust is confined to small radii, the resulting SEDs are quite blue in both type 1 and type 2 orientations of the torus \citep[see][, Figs.~6 \& 9]{Hon10}. On the other hand, more shallow dust distributions ($a=-0.5\ldots-1.0$) show redder colors for all orientations. Thus, in the framework of a clumpy torus, the similarity in mid-IR spectral indices can be naturally explained. A larger difference in spectral indices, however, may emerge at wavelengths shorter than 8\,$\,micron$.

\subsection{The silicate feature and its relation to AGN properties}\label{sec:si}

In Sect~\ref{sec:siobs} we discussed qualitatively the general appearance of silicate features at high spatial resolution. Here, we aim for a quantitative analysis of the silicate feature and its relation to other observed properties. For that, we fit a linear continuum to the 8.5\,$\micron$ and 12.5\,$\micron$ flux of each object. From that we interpolate the expected linear continuum flux $F_\mathrm{c}$ at 9.8\,$\micron$ around the center of the silicate feature (but see Sec.~\ref{sec:siobs} for details on the central wavelength). Then, we take the ratio between observed flux and fitted continuum flux, $F_\mathrm{Si}/F_\mathrm{c}$, as a measure for the strength of the silicate feature, which is similar to the strategy used by \citet{Hao07}. The silicate feature strength of each object is listed in Table~\ref{tab:specprop}.

Fig.~\ref{fig:siprop} shows the silicate strength plotted against several observed parameters. The right axis shows the translation into an observed optical depth, $\tau_\mathrm{silicate}$, within the silicate feature. It has to be pointed out, however, that this does not reflect the actual optical depth along the line-of-sight towards the central engine, but is a result of the radiative transfer within the torus \citep[see also][]{Hon10}. In the left panel, we show the silicate strength as a function of mid-IR luminosity $\Lm$. As in all plots, the type 1 AGN are marked by blue triangles, and red squares are used for type 2 AGN. For reference, we put NGC~1068 in all of the plots based on the Gemini spectrum of the nucleus reported in \citet{Mas06}. The type 1 AGN are all clustered around $\log F_\mathrm{Si}/F_\mathrm{c} = 0$, reflecting the absence or extremely shallow silicate features. They do not show any dependence of the silicate feature on $\Lm$ in the covered luminosity, with a nominal relation $\log F_\mathrm{Si}/F_\mathrm{c} \propto (0.00\pm0.03)\times\log\Lm$ (dotted line in the left panel of Fig.~\ref{fig:siprop}). The type 2 AGN show a much larger variety of silicate feature strengths from moderate absorption to slight emission, as seen also in Fig.~\ref{fig:VISIR_type2}. The deepest feature is observed in NGC~7582. This galaxy shows a well-known star-burst ring close to the nucleus which is at high inclination \citep[e.g.][]{Wol06}. It seems quite likely that a large fraction of the obscuration in this object is not intrinsic to the torus but caused by dust located in or around the star-burst ring or the inner host galaxy. In general the silicate absorption features seem to become shallower with luminosity in the type 2 AGN. The dashed line in the left panel of Fig.~\ref{fig:siprop} is the nominal fit to the type 2 sub-sample, $\log F_\mathrm{Si}/F_\mathrm{c} \propto (0.12\pm0.06)\times\log\Lm$. We would not call this a correlation (Spearman rank $\rho=0.58$, null-hypothesis probability $8.2\times10^{-2}$), but the data suggest at least that the silicate feature shows stronger differences between type 1 and type 2 AGN at lower than at higher luminosities. If this is really true, it indicates that the dust distribution properties such as the radial dust distribution or the average number of clouds along the equatorial line-of-sight \citep[see][for more details]{Hon10} may change with luminosity. This would go in line with the possible trend of bluer spectral indices for higher luminosity objects mentioned in Sect.~\ref{sec:specind}. Based on the discussions in \citet{Hon10}, it is possible that higher luminosity AGN either have a more compact dust distribution \citep[see also][]{Pol08} or have less obscuring clouds in the torus \citep[e.g. see][]{Hon07b}. Follow-up studies on a much larger sample and luminosity range should be done to further investigate these trends. We will continue this discussion in Sect.~\ref{sec:lowhighL}.

A fundamental prediction of all kind of torus models is that line-of-sight obscuration, spectral color, and silicate feature are correlated. As a first-order approximation, we would expect that the SEDs of type 2 AGN are slightly redder than type 1 AGN. If we have a face-on view onto the torus, the near-to-mid-IR SED is dominated by hot-dust emission from the inner part of the torus. In type 2 AGN, the inner hot dust is obscured by cooler dust at larger distances from the AGN. This difference in observed dust temperature should lead to differences in the spectral color (but see Sect.~\ref{sec:specind}). On the other hand, since hot dust is expected to show the silicate feature in emission and cold (optically-thick) dust to show a silicate absorption feature, spectral index $\alpha$ and silicate feature strength $F_\mathrm{Si}/F_\mathrm{c}$ are supposed to be correlated. In the middle panel of Fig.~\ref{fig:siprop} we plot $\log F_\mathrm{Si}/F_\mathrm{c}$ against the mid-IR flux ratio $\log F_\nu(12.5\,\micron)/F_\nu(8.5\,\micron)$ and spectral index $\alpha$. The type 1 AGN do not show a dependence of the silicate feature strength on the spectral index while a range of $\alpha=-0.8\ldots\,-2.2$ is covered. Apparently, the different spectral indices are not an effect of obscuration (assuming that the silicate feature depth is tracing the line-of-sight obscuration to some degree since the tori are seen face-on). Instead it may be a signpost of different dust distributions. We showed in \citet{Hon10} that the spectral slope in the mid-IR is quite sensitive to the power law index $a$ of the radial dust distribution $\eta_r\propto r^a$ where $r$ is the distance from the AGN. The steeper the distribution (i.e. dust more concentrated to the inner part of the torus), the bluer the mid-IR SED. On the other hand, if the dust distribution is shallower, the mid-IR is redder. Based on these models, we would conclude that the type 1 AGN in our sample show some variety in their radial dust distribution, becoming steeper from right to left in this plot. We will pick-up this suggestion when modeling our data in Sect.~\ref{sec:model}.

The type 2 AGN show some weak trend of redder colors with deeper silicate absorption features. The only exception is NGC~7582 were the silicate feature is much deeper for the given $\alpha$ than in the rest of the objects. Again, this might be connected to additional absorption outside the torus (see above). The nominal relation between silicate feature strength and spectral steepness for type 2s is $\log F_\mathrm{Si}/F_\mathrm{c} \propto (-0.85\pm0.20)\,\times\,\log F_\nu(12.5\,\micron)/F_\nu(8.5\,\micron)$ (excluding NGC~7582), or expressed as spectral index, $\log F_\mathrm{Si}/F_\mathrm{c} \propto (0.14\pm0.03)\,\cdot\,\alpha$ (Spearman rank $\rho=-0.88$, null-hypothesis probability $1.6\times10^{-3}$). This result is, at least qualitatively, in agreement with what we would expect from a clumpy dust torus around the AGN. Finally, we can test how well the observed obscuration properties in X-ray and the mid-IR are related. In the right panel of Fig.~\ref{fig:siprop}, we plot the observed silicate feature strength $F_\mathrm{Si}/F_\mathrm{c}$ in the mid-IR against the Hydrogen column densities $\nh$ observed in the X-rays. Despite some scatter, we find a correlation $F_\mathrm{Si}/F_\mathrm{c} \propto (0.12\pm0.03)\,\times\,\log\nh$ (Spearman rank $-0.71$, null-hypothesis probability $9.9\times10^{-4}$) which is shown as a dashed line in Fig.~\ref{fig:siprop}. This is consistent with the analysis of \citet{Shi06} who used \Spitzer data at lower spatial resolution. $F_\mathrm{Si}/F_\mathrm{c}$ can be converted into an \textit{observed} optical depth $\tau_\mathrm{silicate}$ in the silicate feature, and we define $F_\mathrm{Si}/F_\mathrm{c}=\exp\left(-\tau_\mathrm{silicate}\right)$ (see right $y$-axis in Fig.~\ref{fig:siprop} and data in Table~\ref{tab:specprop}). In this way, the correlation between observed silicate feature depth and Hydrogen column density becomes
\begin{equation}
\tau_\mathrm{silicate} = (0.00\pm0.73) + (0.14\pm0.03)\,\times\,\log N_\mathrm{H;22}
\end{equation}
where $N_\mathrm{H;22}$ is $\nh$ in units of $10^{22}$\,cm$^{-2}$. The observed scatter in this correlation may be a signpost of a clumpy torus where type 1 and type 2 AGN are less differentiated. We remind again that the \textit{observed} $\tau_\mathrm{silicate}$ is not the same as the \textit{actual} optical depth along the line-of-sight but reflects the complicated radiative transfer inside the torus. In fact, based on clumpy torus models we would expect that the \textit{actual} optical depth along the line-of-sight is much stronger correlated with $\nh$ unless a sizable fraction of the Hydrogen column is located inside the sublimation radius (e.g. within the broad-line region). Note that NGC~7582 has been excluded from this analysis due to the concerns mentioned above.

Based on the discussion in Sect.~\ref{sec:siobs}, it is possible that the silicate emission features have a different peak wavelength than the center of the silicate absorption features. This might potentially change the outcome of our analysis on the silicate feature strength relations. In order to investigate this possibility, we extracted $F_\mathrm{Si}/F_\mathrm{c}$ at 10.3\,$\micron$ instead of 9.8\,$\micron$ for the type 1 sub-sample. The resulting feature strengths are plotted as gray circles in the right panel of Fig.~\ref{fig:siprop}. Changes to the original analysis are only moderate and the resulting correlation is shown as a gray-dotted line. The results remain consistent within error bars (here, $F_\mathrm{Si}/F_\mathrm{c} \propto (0.14\pm0.03)\,\times\,\log\nh$, Spearman rank $\rho=-0.74$, null-hypothesis probability $4.5\times10^{-4}$). Thus, our general analysis does not suffer significantly from any possible shift of the central wavelength in silicate emission features.

\begin{figure*}
%\sidecaption
\centering
\includegraphics[width=1.0\textwidth]{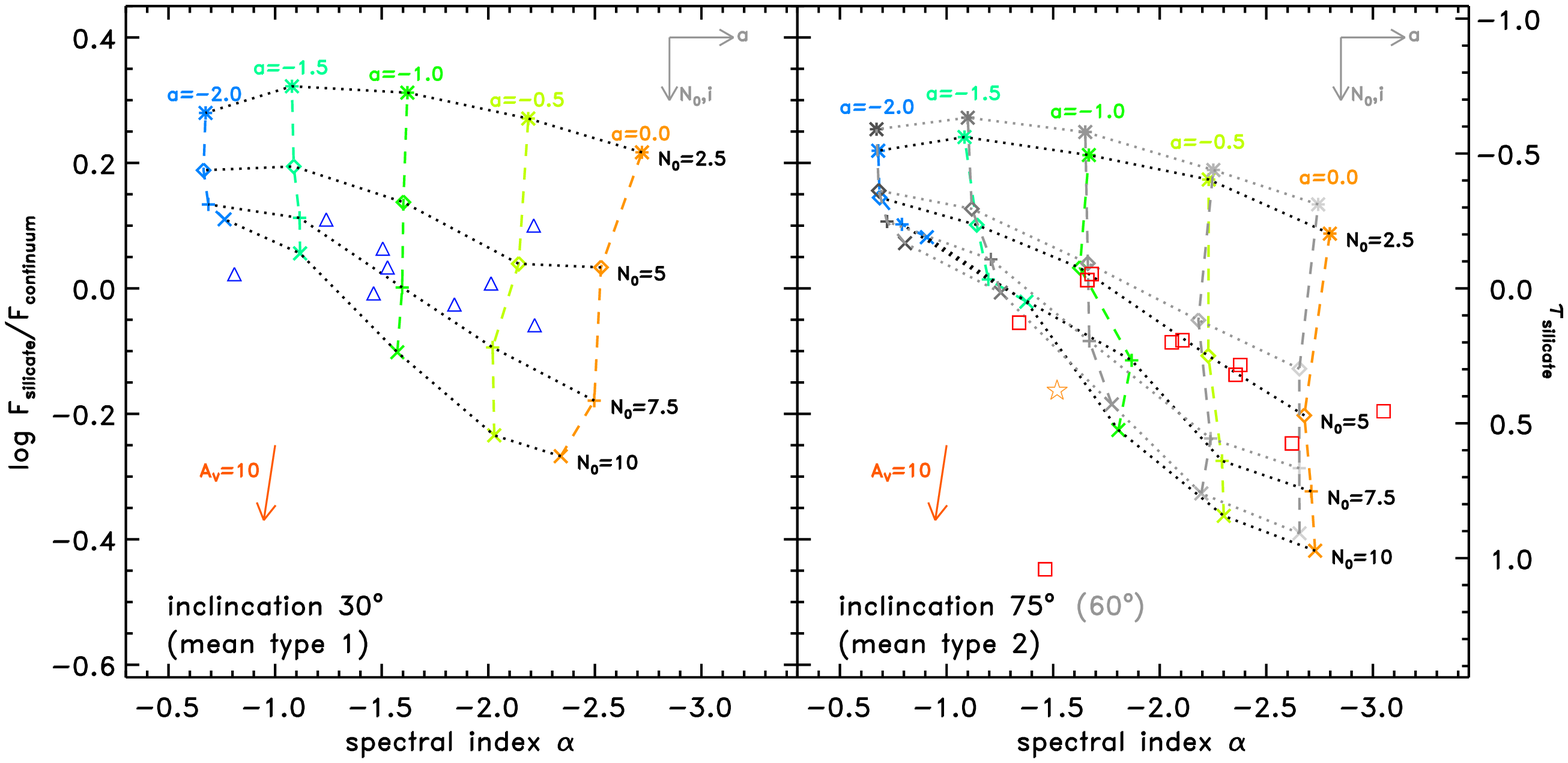}
\caption{Comparison of observed mid-IR properties of our type 1 and type 2 AGN to model SEDs simulated with our 3d clumpy torus model. \textit{Left}: $F_\mathrm{Si}/F_\mathrm{c}$ and $\tau_\mathrm{silicate}$ of the type 1 sub-sample (blue triangles) plotted against the spectral spectral index $\alpha$ of the mid-IR continuum emission. Overplotted are model predictions for parameters ranging from $a=0.0$ (orange) to $a=-2.0$ (light blue) and $\No=2.5$ (asterisks) to $\No=10$ (x-shapes) for a mean type 1 inclination of $i=30^\circ$. The black arrows note the directions in which the model points change when varying $a$, $\No$, and $i$. \textit{Right:} $F_\mathrm{Si}/F_\mathrm{c}$ and $\tau_\mathrm{silicate}$ of the type 2 sub-sample (red squares) plotted against the spectral spectral index $\alpha$ of the mid-IR continuum emission. Overplotted are model predictions for parameters ranging from $a=0.0$ (orange) to $a=-2.0$ (light blue) and $\No=2.5$ (asterisks) to $\No=10$ (x-shapes) for a mean type 2 inclination of $i=75^\circ$ (colored symbols) and $i=60^\circ$ (gray symbols, illustrating the inclination effect). The gray arrows are a rough illustration as a guide for the reader in which direction the model points mostly change when increasing $a$, $\No$, and $i$. For reference, NGC~1068 is shown as an orange star, based on the mid-IR spectrum from \citet{Mas06}.}\label{fig:model}
\end{figure*}

\section{Modeling the data with 3D clumpy torus models}\label{sec:model}

\subsection{Constraining torus parameters from VISIR N-band spectro-photometry of nearby AGN}

In \citet{Hon10}, we prepare interpretation of mid-IR observations of AGN using our 3D clumpy torus model \citep{Hon06}. General dependencies of SEDs and interferometric visibilities on different model parameters have been discussed and it has been shown that the mid-IR SEDs are most sensitive to (1) the radial distribution of dust clouds which we parametrize as a power-law $\eta_r(r) \propto r^a$ \citep[for details please see Sect.~2.5 in][]{Hon10} and (2) the obscuration inside the torus. The torus-internal obscuration is parametrized by the average number of clouds, $\No$, along an \textit{equatorial} line-of-sight. We remind the reader that $\No$ is \textit{not} equal to the number of clouds along the \textit{actual} line-of-sight unless the torus is seen edge-on. We argued that SEDs may serve as a tool to constrain $a$ and maybe $\No$ if the strength of the silicate feature and the spectral slope in the mid-IR are simultaneously taken into account.

The mid-IR data presented in this paper has a quite small wavelength coverage from approximately $8-13\,\micron$. This limits the possibilities of modeling IR SEDs, e.g. as recently done for NGC~1068 \citep{Hon08b}. As mentioned in the previous section, the two main observational parameters that we can derive from our VISIR data are the spectral index $\alpha$ of the mid-IR continuum and the depth of the silicate feature $F_\mathrm{Si}/F_\mathrm{c}$ (or translated into an observed optical depth $\tau_\mathrm{silicate}$). In the middle panel of Fig.~\ref{fig:siprop} we plotted both parameters against each other and discussed the observed properties of the type 1 and type 2 sub-samples. Now, we want to compare these observed mid-IR properties to our torus models and see if we can constrain parameters. For that we simulated a model grid varying $a$ and $\No$ for inclinations from $i=0^\circ$ (pole-on) to $90^\circ$. As discussed in \citet{Hon10}, other model parameters do not have a significant influence on the SED, except for the half-opening angle $\theta_0$ of the torus, which we discuss below. For the moment, we use the common assumption of $\theta_0=45^\circ$.

In Fig.~\ref{fig:model} we show model predictions for the silicate feature strength $F_\mathrm{Si}/F_\mathrm{c}$ (or observed optical depth $\tau_\mathrm{silicate}$ in the silicate feature) and the spectral index $\alpha$ of the mid-IR continuum. In the left panel, we show simulations for an inclination angle of $i=30^\circ$ which reflects a typical type 1 AGN line-of-sight for a half opening angle of $\theta_0=45^\circ$. Model results for different radial power law indices are color coded from $a=-2.0$ (blue) to $a=0.0$ (orange). For each $a$, several values for $\No$ have been calculated and are marked by different symbols in the plot from $\No=2.5$ (asterisks) to $\No=10$ (x-shapes). Dashed-colored lines connect symbols with the same $a$ but varying $\No$, while dotted-black lines represent the same $\No$ for different $a$ values. The right panel of Fig.~\ref{fig:model} shows the same model grid but for a typical type 2 inclination of $i=75^\circ$. We also overlaid the grid for $i=60^\circ$ to illustrate how inclination effects change the results. In general, changing $a$ has a strong effect on the spectral index $\alpha$ \citep[as discussed in][]{Hon10} and some effect on $F_\mathrm{Si}/F_\mathrm{c}$. On the other hand, varying $\No$ and $i$ has almost no effect on $\alpha$ but changes $F_\mathrm{Si}/F_\mathrm{c}$. It is noteworthy that the spectral index $\alpha$ is very insensitive to the torus inclination, even when changing from edge-on to face-on line-of-sights. As discussed, this is characteristic for clumpy models. The gray arrows in Fig.~\ref{fig:model} are intended to help the reader, summarizing the rough trends of how the different parameters change the place of an object in the model grid.

The model grids in Fig.~\ref{fig:model} illustrate that some fundamental parameters of the dust torus can be constrained even using the limited wavelength range covered by our observations. We overplot the type 1 and type 2 sub-samples in the left and right panel of Fig.~\ref{fig:model}, respectively, in order to see what region in parameter space is occupied. As already discussed, the most robust constraint can be put on $a$ since it is the almost exclusive parameter to determine an object's place in horizontal direction. The type 1 AGN in the left panel populate the range between $a=0.0$ to $a=-1.5$, with some clumping roughly at $-1.0\pm0.5$. ESO~323--G77 is the only exception being located between $a=-1.5$ and $a=-2.0$ but outside the displayed model grid. It is discussed in more detail below. Interestingly, most type 1s also cover a rather narrow range in vertical direction, somewhere between $\No=5$ and $7.5$, and they follow approximately a line of equal $\No$ for varying $a$. NGC~7213 is slightly above the other objects, suggesting a lower value for $\No$. Since it is not a classical Seyfert galaxy but a LINER, the offset from the Seyfert 1s might indicate some intrinsic difference in the dust structure. On the other hand, it may also result from an inclination effect: If NGC~7213 is seen more pole on than the other objects, then we would expect that for the same $a$ and $\No$ the object is placed slightly above the others (see also inclination effect illustrated in the right panel of Fig.~\ref{fig:model}). However, at low inclinations, any viewing-angle effect is small and not able to explain all of the offset of NGC~7213.

The right panel of Fig.~\ref{fig:model} shows the type 2 AGN of our sample. Most objects are again located between $a=0.0$ and $a=-1.5$ and they follow the line of constant $\No=5$ for inclination $i=75^\circ$ or $\No\sim6$ for $i=60^\circ$. There are, however, a number of outliers from this plot, most notably NGC~7582. For this galaxy we already remarked that additional extinction from a circumnuclear star-burst ring might interfere with the emission from the nucleus. Consequently, the observed mid-IR properties are not generic properties of the torus emission. To study the possibility if such additional (screen) absorber may be responsible for outlier in the plot, we used the extinction curves of our dust composition \citep[ISM dust with Ossenkopf silicates; see][for details]{Hon10} and derived the change of mid-IR spectral properties for $A_V=10$. The resulting shift is shown as an orange arrow in both panels of Fig.~\ref{fig:model}. Actually NGC~7582 and MCG--5--23--16 -- the objects in our sample which are located below the model grid -- have a comparably high inclination of the host-galaxy which may contribute some extinction screen to the nuclear emission. On the other hand, MCG--5--23--16 is very close to the model grid and the offset can be easily explained by either a torus inclination $i>75^\circ$ or a torus with $\No=10$ or slightly higher. Please note that ``reddening'' (extinction) is actually making the $N$-band mid-IR colors slightly bluer, which is an effect of the extinction curve at these wavelengths. Another interesting object is NGC~5643. This low-luminosity AGN has the reddest colors in our sample. It follows quite well the line of constant $\No$ occupied by the other AGN, but is placed at $a>0.0$. This would imply that the torus is much more extended than in the other objects, compared to the intrinsic scale, something that might be tested by IR interferometry.

Finally, we will discuss ESO~323--G77 which is an outlier in the type 1 plot (left panel of Fig.~\ref{fig:model}). It is located slightly below the model grid between $a=-1.5$ and $-2.0$. Based on our modeling, the offset cannot be explained by either an inclination effect or a higher $\No$. Inclination is excluded since ESO~323--G77 is even below the type 2 AGN model grid in the left panel of Fig.~\ref{fig:model}. Moreover, for such steep dust distributions, most clouds are located in the innermost part of the torus close to the sublimation radius. Adding more clouds to the torus will dramatically increase the torus-internal obscuration and moves the models rather to the right than downward in the plot. This effect can already be seen in the model grid for $a=-2.0$ and $\No=10$ and becomes even stronger for higher $\No$. On the other hand, ESO~323--G77 is showing some absorption pattern in the optical and the X-rays (see Table~\ref{tab:scales}) which might result in a down-left drift in the plot (see orange arrow). It is located in the same area of the $\LmLx$-correlation as MCG--3--34--64 and IC~5063 for which we found extended emission probably coming from the NLR. Since ESO~323--G77 is a type 1 AGN with a more or less face-on line-of-sight, dust in the NLR can cause screen extinction which may explain the offset position in Fig.~\ref{fig:model}. In this case it is reasonable to conclude from Fig.~\ref{fig:model} that the intrinsic dust distribution power-law index of ESO~323--G77 is most likely somewhere between $a=-1.5$ and $-2.0$ as expected from the relatively blue mid-IR color.

In summary, based on our modeling we find that the Seyfert galaxies in our sample have a torus with a rather shallow radial dust distribution with a typical power law index of approximately $a=-1\pm0.5$. This parameter is quite solidly constrained by the mid-IR spectral indices. For the presented model grid, we also find a typical range for the average number of clouds along an equatorial line-of-sight of $\No\sim 5-8$ for the whole sample. It is, however, difficult to pin down this parameter for each individual object since inclination effects can slightly alter the results. As of yet, we did not discuss how the vertical structure of the torus (represented by the half-opening angle, $\theta_0$, of the torus) influences the models. In fact, $\theta_0$ does not have an effect on the horizontal position of the object (exclusively determined by $a$) but can change its vertical position, i.e. the strength of the silicate feature. We find that increasing $\theta_0$ from $45^\circ$ to $60^\circ$ has a similar effect as decreasing the inclination angle by about $15^\circ$. If we consider $\theta_0=45^\circ$ as a reasonable \textit{sample average}, the range of $\No\sim5-8$ found as a typical value for the whole sample is still valid, although it might be complicated to constrain $\No$, $\theta_0$, and $i$ for each \textit{individual} object.

\begin{figure}
%\centering
\includegraphics[width=0.5\textwidth]{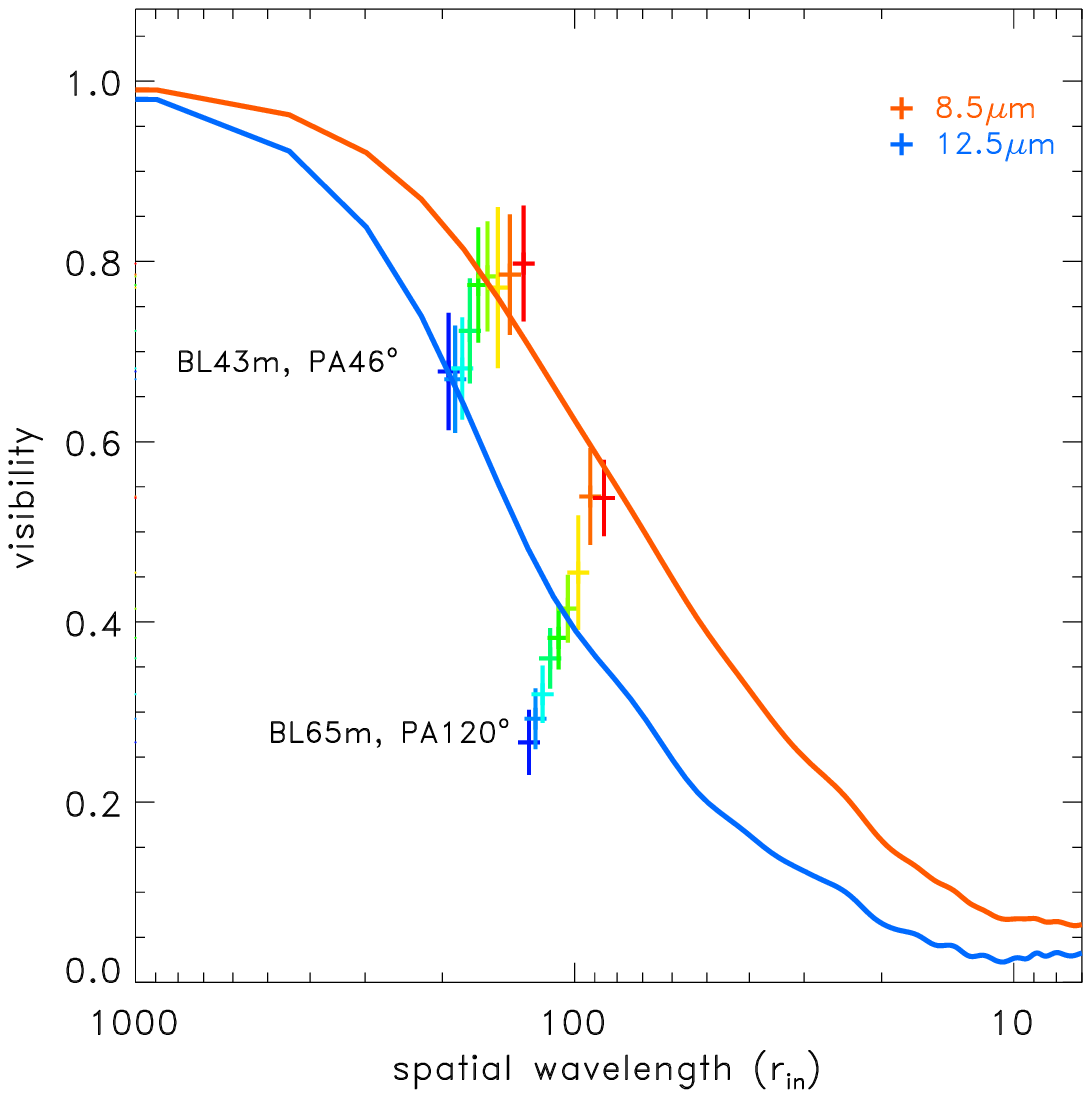}
\caption{Comparison of the VLTI/MIDI $8-13\,\micron$ interferometry of NGC~3783 with our 3D clumpy torus models. The interferometry data at projected baseline length 65\,m have been reported by \citet{Bec08}, the 43\,m data was presented in \citet{Kis09}. We plot the observed $8-13\,\micron$ visibility (color-coded from red to blue) against the spatial wavelength in units of the inner radius $r_\mathrm{in}$ \citep[see][for details]{Kis09}. Overplotted are the model visibilities for 8.5\,$\micron$ (red line) and 12.5\,$\micron$ (blue line) based on model parameters derived from Fig.~\ref{fig:model} ($a=-0.75$, $\No=7$). The 8.5 and 12.5\,$\micron$ data have the same colors as the corresponding model.}\label{fig:N3783vis}
\end{figure}

\subsection{Comparison to IR interferometry}\label{sec:intf}

While our modeling approach based on the spectral slope and silicate feature strength is very limited in wavelength range, we can test its results and predictions. For that, we compare the derived radial dust distribution to recent results from IR interferometry. In \citet{Kis09}, we argued that IR interferometry of type 1 AGN can be used to determine the surface brightness profile of the dust torus. For that, the $N$-band visibilities observed by VLTI/MIDI have been plotted against an object-intrinsic spatial scale (the near-IR reverberation radius, which is presumably indicative for the inner boundary of the dust distribution). It has been found that the brightness profile in NGC~3783 depends on the radius $\propto r^{-2}$. We can now try to reproduce this finding by our models using model parameters as constrained by Fig.~\ref{fig:model}. For NGC~3783 we find that $a\sim -0.75$ and $\No\sim7$ for $i=30$. Based on these parameters, we simulated images and calculated corresponding mid-IR visibilities. The results at $8.5\,\micron$ and $12.5\,\micron$ are shown in Fig.~\ref{fig:N3783vis}. The model visibilities are in general agreement with the available data, despite some overestimation of the visibility at longer wavelengths for the long projected baseline (baseline length 65\,m, PA~120$^\circ$). The fact that the observed visibilities are lower in PA~120$^\circ$ may be due to some elongation of the source. Based on our clumpy torus models we would not expect that the torus itself produces these kind of deviation. Interestingly the larger size in PA~120$^\circ$ is pointing into a similar direction as the polarization angle \citep[$\theta_\mathrm{pol}\sim135^\circ$][]{Smi02} which implies that the mid-IR nucleus is elongated \textit{towards the NLR}. This may place NGC~3783 into the same group of objects as IC~5063 and MCG--3--34--64 which show some contribution from the NLR to the mid-IR emission (although for NGC~3783 on much smaller scale and fraction), in particular to longer wavelengths. We leave a more detailed analysis of the interferometric measurements to a much larger dataset that was obtained recently (H\"onig et al. in preparation). Despite these details, NGC~3783 as a case study illustrates that the brightness profiles of the dust distribution are strongly connected with the mid-IR slope $\alpha$, and both are very sensitive to the radial distribution of the dust.

In the left panel of Fig.~\ref{fig:model}, we also show NGC~1068 as an orange asterisk. It is located slightly below the $i=75^\circ$ model grid in the range of $a=-1.0\ldots-1.5$ and $\No\sim10$. Recently, we presented detailed modeling of the full IR SED of NGC~1068 at high-spatial resolution simultaneously with near- and mid-IR interferometry \citep{Hon06,Hon07a,Hon08b}. Since IR interferometry resolves the spatial brightness profile, it is considered as very constraining for model parameters. For NGC~1068 we found typical parameters of $a\sim-1.5$, $i=70-90^\circ$, and $\No\sim10$ which is in excellent agreement with the analysis presented here based on the mid-IR spectral index and the silicate feature strength. 

In summary we find that the results from our analysis of the mid-IR spectral index $\alpha$ and the silicate feature strength is in good agreement with results from IR interferometry. Model parameters derived by our spectro-photometry approach are consistent with a much more detailed analysis of the IR emission of NGC~1068 and can reproduce interferometric measurements of NGC~3783. In turn, based on the analysis of the mid-IR spectra, we can make predictions about the spatial brightness profiles of AGN for interferometry. For instance, we expect that ESO~323--G77 and MARK~509 have quite compact profiles (i.e. high visibility) while objects like NGC~3227, NGC~7213, or NGC~7469 have rather extended profiles (i.e. low visibilities) when scaled for their intrinsic radii.

\subsection{A luminosity dependence of the torus properties?}\label{sec:lowhighL}

One interesting aspect of the torus is whether or not there are any differences in its structure at low and high AGN luminosity. Obscuration statistics of type 1 and type 2 AGN suggest that the relative fraction of type 1 AGN increases with luminosity \citep[e.g.][]{Sim05,Mai07,Has08}. This phenomenon is commonly explained by a decrease of torus covering factor with luminosity -- and referred to as the ``receding torus''. On the other hand, if the covering factor decreases, the ``reprocessing ratio'' $\Lm/\Lx$ ratio is also expected to decrease since less torus surface is heated by the AGN. However, based on the $\LmLx$-correlation by \citet{Gan09}, we find that $\Lm/\Lx \propto \Lx^{0.11\pm0.07}$ which means that the reprocessing ratio does not significantly change with luminosity (maybe marginally increases; but we note that when using the bolometric luminosity instead of $\Lx$, the ratio might slightly decrease with luminosity, depending on the bolometric correction used; see also \citet{Gan09}, Sect.~4.5). In \citet{Hon07b}, we suggested a possibility to explain these observations: increasing AGN radiation pressure may cause large dust clouds to be driven out of the torus at higher luminosities. Consequently, the absorbing column towards the AGN decreases and it becomes less likely that the line-of-sight towards the nucleus is obscured. On the other hand, the covering factor and reprocessing ratio will not change.

The question remains, however, if there are direct hints that the torus \textit{dust distribution} changes with luminosity. The most direct way of probing this possibility is, of course, IR interferometry. On the other hand, due to the sensitivity of current interferometers, one is limited to few objects and the luminosity coverage is limited. In the previous sections we showed that the radial structure of the torus and the torus-internal obscuration properties can be constrained by simultaneously comparing the mid-IR spectral index $\alpha$ and the silicate feature strength. In particular, constraining the radial power-law index $a$ using $\alpha$ seems to be quite solid (see also Sect.~\ref{sec:intf}).

The AGN in our sample have typical luminosities of nearby Seyfert galaxies, i.e. in the range of $10^{42}\,\ergs$ to $10^{45}\,\ergs$, and conclusions drawn about the typical torus properties of $a=-1\pm0.5$ and $\No=5-8$ are only valid for these luminosities. Recently, \citet{Pol08} presented rest-frame near-to-mid-IR SEDs of 23 type 2 QSOs with luminosities between $10^{46}\,\ergs$ to several $10^{47}\,\ergs$. Interestingly, most of them showed quite blue IR spectral indices. Using the clumpy torus models, it was concluded that the typical radial power-law indices in the type 2 QSOs were tracing a quite steep dust distribution, with $a\sim-2\ldots-3$. If this is indeed a typical value for luminous AGN, it is much steeper than what we find in our intermediate-luminosity sample. Consequently, there could be a change of the torus structure with luminosity.

At the low-luminosity end, we do not have a good coverage of mid-IR data where spatial resolution is sufficient to isolate the torus emission. However, we can compare the lower-luminosity objects in our sample to those at the higher luminosity end. In fact, in Sect.~\ref{sec:specind} we note that the reddest objects in both type 1 and type 2 sub-samples are those with the lowest luminosities (NGC~3227 and NGC~7213 for type 1s, NGC~5643 and ESO~428--G14 for the type 2s). This would suggest that the lower-luminosity AGN have much more shallow dust distributions, i.e. the torus is relatively more extended as compared to higher-luminosity objects. If true, we might see an imprint of this behavior in the mid-IR size-luminosity relation, e.g. as it can be constructed from interferometry. The prediction would be that for a sufficiently large sample the relation does not scale as $r \propto L^{1/2}$ but with a slightly smaller power-law index. Similarly, near-IR interferometric sizes would be systematically larger than IR reverberation-mapping radii at lower luminosities, and would better match each other at higher luminosities \citep[cf.][for Keck interferometry results of 4 objects]{Kis09b}.

\section{Summary}\label{sec:summary}

In this paper, we presented high-spatial resolution mid-IR spectro-photometry of a sample of nearby AGN. Based on the selection criteria, the sample cannot be considered complete but the objects are typical for local intermediate-luminous AGN, i.e. Seyfert 1 and Seyfert 2 galaxies ranging from Compton-thin to Compton-thick X-ray obscuration. The ground-based mid-IR data was taken with the VLT mid-IR imager and spectrograph VISIR, and it represents, to our knowledge, the largest sample of mid-IR $N$-band spectra of AGN with angular resolution $\le$0$\farcs$4. The main characteristics of these data are:
\begin{list}{$\bullet$}{}
\item Our ground-based observations isolate the nuclear dust emission of the torus from host-galactic sources. PAH emission features, commonly associated with star-formation in the host galaxy, are mostly absent in our data (i.e. at spatial scales $<$100\,pc), although seen in \Spitzer data at lower spatial resolution. We showed that subtracting the nuclear VISIR emission from the \Spitzer data reveals spectra reminiscent of a star-formation template.   
\item The mid-IR spectra show only moderate silicate features. While silicate absorption is easily identified in type 2 AGN, the corresponding emission feature in type 1 AGN is very weak, if present at all. Thus, the weakness of the silicate feature is an intrinsic property of the emitting region which is $<$25--160\,pc, depending on the object. This points to an origin in the circumnuclear dust distribution -- the ``dust torus''.
\item The strongest silicate emission features are seen in the type 2 AGN NGC~2110 \citep[see also][]{Mas09} and the broad-line LINER NGC~7213. We analyzed the silicate emission feature of the latter object and found a shift of central wavelength of the silicate feature to $\sim$10.5$\,\micron$. This shift is probably not exclusively caused by radiative transfer effects due to absorption as recently suggested, but might indicate different dust composition than in the standard ISM, or a radial change of grain size and/or chemistry in the dust distribution.
\end{list}

The observed mid-IR luminosities have been compared to other AGN luminosity tracers, namely the 2$-$10\,keV luminosity $\Lx$ and the luminosity of the [\ion{O}{iii}]($\lambda5007\,\AA$) narrow emission line. We showed that correlations of both tracers with $\Lm$ are quite strong and tighter than between $\Lx$ and $\Loiii$. It may be worthwhile considering $\Lm$ as an isotropic AGN luminosity tracer. 

Three object in our sample, MCG--3--34--64, ESO~323--G77, and IC~5063, are outliers in the otherwise tight $\LmLx$-correlation \citep{Hor08,Gan09}. We evaluated mid-IR images in 4 different filters per object and found that they show significant extension and elongation. In IC~5063, the elongation direction is consistent with HST [\ion{O}{iii}] emission maps. We suggest that a sizable fraction of the mid-IR emission in these objects is originating in the narrow-line region, possibly caused by dust located there, similar to the mid-IR-bright NLR in NGC~1068. This additional emission may be the reason for the mid-IR ``overluminosity'' in the $\LmLx$-correlation.

Since the wavelength range covered by our data is limited to $8-13\,\micron$, the main properties that we can use for analyzing the dust emission are (1) the mid-IR continuum spectral index $\alpha$, and (2) the strength of the silicate feature. We found that there are at best mild trends of increasing $\alpha$ with increasing $\Lm$ and increasing $\alpha$ with increasing Hydrogen column density $\nh$. The weak trend of the $\alpha$ vs. $\nh$ shows that strong or weak obscuration does not have a significant impact on the overall color of the mid-IR SED, consistent with the as yet not detected difference between the $\LmLx$-correlation of type 1 and type 2 AGN \citep[see also][]{Gan09}. As suggested in \citet{Hon10}, the distribution of the dust around the AGN has much more impact on the mid-IR spectral slope than obscuration or orientation effects. We also compared the strength of the silicate feature $F_\mathrm{Si}/F_\mathrm{c}$ (or observed $\tau_\mathrm{silicate}$) to the mid-IR luminosity and the Hydrogen column density. We find a similar correlation of silicate feature strength and $\nh$ as \citet{Shi06}, corresponding to $\tau_\mathrm{silicate} \propto 0.14\,\times\,\log\nh$. Interestingly, there might be a difference between silicate feature strength at lower and higher luminosity in our sample: type 1 and type 2 objects at lower $\Lm$ seem to show a stronger difference in silicate features than objects at higher $\Lm$ where type 1 and type 2 features appear rather similar. A larger sample of objects with high angular resolution is required to see if the trend can be confirmed. 

Finally, we plot silicate feature strength against mid-IR spectral index $\alpha$. Such an analysis is motivated by our 3D clumpy torus model where we found that simultaneously accounting for silicate feature and spectral slope may be very constraining for torus parameters. We overplot the observed properties of type 1 and type 2 AGN with predictions from our model, varying (1) the radial dust distribution power law index $a$ (radial distribution $\eta_r\propto r^a$), (2) the average number of clouds $\No$ along the line-of-sight in the equatorial plane, and (3) the torus inclination $i$. Our main conclusions are:
\begin{list}{$\bullet$}{}
\item $\alpha$ versus silicate-feature-strength plots are quite constraining for torus parameters since $\alpha$ is almost exclusively sensitive to $a$ and the silicate strength is mostly sensitive to $\No$ and $i$. We discussed how the position of an object is depending on these and possible other model parameters and concluded that $a$ can be constrained for each individual object while $\No$ may be better derived as a sample mean.
\item For our Seyfert sample, we find typical values of $a=-1.0\pm0.5$ and $\No=5-8$. The sample average values are similar for type 1 and type 2 AGN, which is a support for the unification scheme. We note, however, that this conclusion is only valid for typical Seyfert AGN as presented in our study.
%, and we cannot exclude the possibility of individual ``real'' type 2s. 
\item Some objects may be suffering from extinction in the host galaxy. In such cases, however, additional screen absorption changes mostly the object's vertical position in the $\alpha$-silicate strength plot.
\item The approach of constraining $a$ and $\No$ based on the mid-IR spectral index $\alpha$ and the silicate features strength has been tested for examples where we also have much broader IR SEDs and/or infrared interferometry (NGC~1068 and NGC~3783). Interferometry directly traces the brightness distribution of the emission region and is most sensitive to the radial distribution of the dust \citep[see][]{Kis09}. We find that the model parameters derived from simultaneous SED and interferometry modeling of NGC~1068 are consistent with parameters derived from the $\alpha$-silicate feature plot. In addition, $\alpha$ and silicate-feature-strength parameters derived for NGC~3783 are able to reproduce the available mid-IR interferometric data. Based on our mid-IR single-telescope data, it is possible to make predictions for interferometry.
\end{list}

Finally, we discuss a possible change of torus properties with luminosity. Based on a comparison with literature data and modeling, we find evidence that the dust tori might be more compact (i.e. $a\approx-2\ldots-3$) for high luminosity AGN and shallower on the low-luminosity end ($a\approx0.0\ldots-0.5$), but we need better selected samples to confirm this trend. If true, this change of the dust distribution in the torus will slightly alter the mid-IR size-luminosity relation and causing differences when comparing near-IR interferometric sizes with near-IR reverberation mapping radii of AGN.

\begin{acknowledgements}
The paper has in part been supported by Deutsche Forschungsgemeinschaft (DFG) in the framework of a research fellowship (``Auslandsstipendium'') for SH. PG is supported by JSPS and RIKEN Foreign postdoctoral fellowships. This research made use of the NASA/IPAC Extragalactic Database (NED) operated by the JPL (Caltech), under contract with NASA.
\end{acknowledgements}

\bibliographystyle{aa}

\end{document}